\documentclass[twocolumn,trackchanges]{aastex63}
\RequirePackage{lineno}
\usepackage{amsmath}
\usepackage{amssymb}
\usepackage{amsthm}
\usepackage{natbib,aasdefs,url,bm}
\usepackage{array}
\usepackage{float}
\usepackage{graphicx}
\usepackage{subfigure}
\usepackage{color}
\usepackage{ctable}
\usepackage{threeparttable} 
\usepackage{CJK}
\usepackage{lineno}
\usepackage{tablefootnote}

\def\hlinewd#1{%
\noalign{\ifnum0=`}\fi\hrule \@height #1 %
\futurelet\reserved@a\@xhline} 

\shorttitle{Structured jet model for jetted TDE AT 2022cmc}
\shortauthors{Yuan, Zhang, Winter, Murase}

\begin{document}
\begin{CJK*}{UTF8}{gbsn}


\title{{Structured Jet} Model for Multiwavelength Observations of the Jetted Tidal Disruption Event \\ AT 2022cmc}

\correspondingauthor{Chengchao Yuan}
\email{chengchao.yuan@desy.de}

\author[0000-0003-0327-6136]{Chengchao Yuan (袁成超)}\affil{Deutsches Elektronen-Synchrotron DESY, 
Platanenallee 6, 15738 Zeuthen, Germany}

\author[0000-0003-2478-333X]{B. Theodore Zhang (张兵)}\affil{Center for Gravitational Physics and Quantum Information, Yukawa Institute for Theoretical Physics, Kyoto University, Kyoto, Kyoto 606-8502, Japan}

\author[0000-0001-7062-0289]{Walter Winter}\affil{Deutsches Elektronen-Synchrotron DESY, 
Platanenallee 6, 15738 Zeuthen, Germany}

\author[0000-0002-5358-5642]{Kohta Murase}
\affil{Department of Physics, Department of Astronomy \& Astrophysics, Center for Multimessenger Astrophysics, Institute for Gravitation and the Cosmos, The Pennsylvania State University, University Park, PA 16802, USA}\affil{Center for Gravitational Physics and Quantum Information, Yukawa Institute for Theoretical Physics, Kyoto University, Kyoto, Kyoto 606-8502, Japan}

\begin{abstract}
AT 2022cmc is a recently documented tidal disruption event (TDE) that exhibits a luminous jet, accompanied by fast-declining X-ray and long-lasting radio/millimeter emission. Motivated by the distinct spectral and temporal signatures between X-ray and radio observations, we propose a multizone model involving relativistic jets with different Lorentz factors. 
We systematically study the evolution of the faster and slower jets in an external density profile, considering the continuous energy injection rate associated with the time-dependent accretion rates before and after the mass fallback time. We investigate time-dependent multiwavelength emission from both the forward shock and reverse shock regions of the fast and slow jets, in a self-consistent manner. 
Our analysis demonstrates that the energy injection rate can significantly impact the jet evolution and subsequently influence the lightcurves. We find that the X-ray spectra and lightcurves could be described by the electron synchrotron emission from the reverse shock of the faster jet, in which the late-time X-ray upper limits, extending to 400 days after the disruption,
could be interpreted as the jet break. Meanwhile, the radio observations can be interpreted as a result of synchrotron emission from the forward shock region of the slower jet. We also discuss prospects for testing the model with current and future observations.
\end{abstract}

\section{Introduction}
Tidal disruption events (TDEs) are astronomical phenomena that occur when a star is torn apart by the gravitational forces of a supermassive black hole as the star orbits in close proximity to the supermassive black hole \citep[SMBH, e.g.,][]{1975Natur.254..295H, 1988Natur.333..523R, 1989ApJ...346L..13E}. The subsequent accretion activities, fueled by the bound mass of the star, can generate a luminous transient spanning a broad electromagnetic spectrum, including radio, optical, ultraviolet, X-ray, and $\gamma$-ray bands. Among the expanding catalog of recorded TDEs, four TDEs, including Swift J1644+57 \citep{2011Sci...333..203B,2011Natur.476..421B}, Swift J2058+05 \citep{2012ApJ...753...77C}, Swift J1112-8238 \citep{2015MNRAS.452.4297B} and AT 2022cmc \citep{2022Natur.612..430A}, have been classified as jetted TDEs. These jetted TDEs display prominent signatures of relativistic outflows, including fast-decaying lightcurves and nonthermal flux intensities in X-ray bands, alongside bright, long-lived radio emission \citep[e.g.,][]{2023NatAs...7...88P, 2024ApJ...965...39Y}.  

It has been thought that the radio emission is explained by synchrotron emission of electrons accelerated at external forward shocks \citep[e.g.,][]{2011Natur.476..425Z,2011MNRAS.416.2102G,2012MNRAS.420.3528M,2012ApJ...748...36B,2013ApJ...767..152Z,2018ApJ...854...86E,2021ApJ...908..125C,2023MNRAS.522.4028M} as the jet propagates into the circumnuclear medium and becomes decelerated, resembling the physics of gamma-ray burst (GRB) afterglows \citep[e.g.,][]{zhang2018physics}. 

{As for X-ray emission from jetted TDEs, interpreting them within the same emission zone for the radio emission} is challenging. Different sites or mechanisms --- such as the magnetically dominated jet \citep{2011Natur.476..421B}, variable accretion near the SMBH horizon \citep{2012Sci...337..949R}, inverse Compton scattering of external photons \citep{2011Sci...333..203B,2016MNRAS.460..396C} --- are proposed to describe the X-ray observations. A self-consistent physical framework is needed to interpret the multiwavelength spectral energy distributions (SEDs) and the temporal evolutions of radio and X-ray emission. 

The recent documented jetted TDE AT 2022cmc at redshift $z=1.19$ provides us another prototype for multiwavelength modeling. AT 2022cmc was discovered by the Zwicky Transient Facility (ZTF) in the optical band \citep{2022Natur.612..430A}. The short variability time scale in the SMBH rest frame, $t_{\rm var}\lesssim1000~{\rm s}/(1+z)$, implies an upper limit of the SMBH mass of few $\times10^7M_\odot$ \citep{2024ApJ...965...39Y}. A relativistic jet with high Lorentz factor $\sim100$ was initially suggested to explain its super luminous and fast evolving X-ray (e.g., $L_X\propto t^{-2}$) and long-lasting radio emissions \citep{2023NatAs...7...88P}. Moreover, an equipartition analysis \citep[e.g.,][]{1998ApJ...499..810C,2013ApJ...772...78B} and a detailed afterglow model reveals the radio/millimeter emitting plasma to be expanding relativistically \citep{2023MNRAS.522.4028M,2024ApJ...965...39Y}, e.g., $\Gamma\lesssim2-5$. 
The two-component jet model with a fast inner component and slow outer component, and has also been exploited to explain the multiwavelength emission from TDEs \citep{2014ApJ...788...32W,2015MNRAS.450.2824M,2015ApJ...798...13L,2023ApJ...957L...9T,2024arXiv240413326S}. Recently, \cite{2024ApJ...963...66Z} demonstrated that the early and late radio emission of AT 2022cmc can be described by the forward shocks of fast and slow jets but in the best-fitting cases the model is insufficient to reproduce X-ray lightcurves, which implies that the radio/millimeter and X-ray may have different origins.

Motivated by the distinct signatures between radio/millimeter and X-ray signals, such as their lightcurves (e.g., long-lasting vs. fast-decaying), their variability time scales (e.g., day-time-scale \citep{2023MNRAS.521..389R} vs. $\sim10^3$~s), and their spectral shapes (e.g., synchrotron self-absorption tail vs. synchrotron broken power-law), we present a multizone model incorporating a fast relativistic jet (narrow outflow, denoted as `fast jet') and a  slow relativistic jet (wide outflow, denoted as `slow jet') capable of explaining the X-ray and radio SEDs and lightcurves simultaneously. {Each jet has a top-hat structure and points towards the observer.} In addition to TDEs, structured jets have been extensively studied in the context of GRBs for a long time \citep[e.g.,][]{2002MNRAS.332..945R,2002ApJ...571..876Z,2021MNRAS.504.5647S,2023arXiv231113671Z}.

{In this work, following the treatment for GRB blast waves \citep{2013MNRAS.433.2107N,zhang2018physics,2023arXiv231113671Z}, we solve the differential equations governing the time evolution of jets sweeping an external medium, taking into account the time-dependent continuous energy/mass injections. We then compute the time-dependent synchrotron and inverse Compton emission from the forward shock and reverse shock regions. Our results demonstrate that the radio/millimeter observations can be explained by the forward shock model of the slow jet, and the fast jet reverse shock synchrotron emissions can reproduce the X-ray spectra and lightcurve. We also argue that the steepening of the late-time (approximately 200$-$400 days after disruption) X-ray lightcurve, as reported in \cite{2024arXiv240410036E}, could be attributed to the jet break as the jet Lorentz factor decreases.}

The paper is organized as follows. In Sec. \ref{sec:accretion_rate}, we model the accretion history. The physical picture and time evolution of jets are presented in Sec. \ref{sec:jet_dyn}. We then apply the dynamics of jets to compute the time-dependent synchrotron and inverse Compton emissions in the jet forward shock and reverse shock regions of the fast and slow jets in Sec. \ref{sec:results}, where the radio/millimeter and X-ray SEDs in three epochs (15-16 d, 25-27 d and 41-46 d in the observer's frame) and lightcurves extending to 400 days after the disruption are also fitted. A discussion and a summary are given in Sec. \ref{sec:discussion} and Sec. \ref{sec:summary}, respectively.

Throughout the paper, we use $T$, $t$, and $t'$ to denote the times measured in the observer's frame, SMBH-rest frame, and jet comoving frame, respectively. The subscripts `f' and `s' will be used to denote the quantities related to fast and slow jets. The notation $Q_{x}$ represents $Q/10^x$ in CGS units unless otherwise specified.

\section{Accretion Rate onto the SMBH}\label{sec:accretion_rate} 
Considering the disruption of a star of mass $M_\star$ and radius $R_\star$ by a SMBH of mass $M_{\rm BH}$, we estimate the tidal radius to be $R_T \approx f_T (M_{\rm BH}/M_\star)^{1/3} R_\star$ \citep[e.g.,][]{1988Natur.333..523R}, where $f_T \sim 0.02$ to $0.3$, accounting for corrections from the stellar internal density profile \citep[e.g.,][]{1989IAUS..136..543P,2015ApJ...806..164P}. For a main sequence star, the radius can be related to the mass via $R_\star=R_\odot(M_\star/M_\odot)^{1-\xi}$, where $R_\odot$ and $M_\odot$ are the solar radius and mass respectively, and the parameter $\xi\sim0.4$ for $1<M_\star/M_\odot<10$ \citep{1990sse..book.....K}. Based on the rather loose constraints on the SMBH mass of AT 2022cmc, e.g., $M_{\rm BH}<5\times10^8M_\odot$ \citep{2022Natur.612..430A}, and the upper limit obtained from the X-ray variability, e.g., $M_{\rm BH}<5\times10^7M_\odot$ \citep{2024ApJ...965...39Y}, we select $M_{\rm BH}=10^7M_{\rm BH,7}M_\odot$ and $M_\star=5M_{\star,0.7}M_\odot$ as the fiducial parameters\footnote{The value of $M_\star$ is degenerate with the energy conversion efficiencies and will be justified in Sec. \ref{sec:discussion}} and obtain the corresponding tidal radius $R_T\simeq1.2\times10^{13}~\rm cm$. After undergoing tidal disruption, approximately half of the stellar material may persist in a gravitationally bound state within an eccentric orbit, ultimately leading to its return and potential accretion onto the SMBH. The fallback time can be estimated using the orbital period of the most tightly bound matter, expressed explicitly as $t_{\rm fb}\approx2\pi\sqrt{a_{\rm min}^3/GM_{\rm BH}}$, where $a_{\rm min}\approx R_T^2/(2R_\star)$ is the semi-major axis of the orbit. In our fiducial case, we have $t_{\rm fb}\simeq3.3\times10^6{~\rm s}~f_{T,-1.2}^{1/2}M_{\rm BH,7}^{1/2}M_{\star,0.7}^{-1/10}$. 

The mass fallback could result in the formation of an accretion disk, with the accretion rate onto the SMBH following a $t^{-5/3}$ proportionality law after the fallback time. Here, $t$ denotes the time measured in the rest frame of the SMBH, which can be correlated with the observation time, $T_{\rm obs}$, through $t=T_{\rm obs}/(1+z)$. Accretion may also occur prior to $t_{\rm fb}$. 
In this study, for the general purposes, we presume a power-law decay in the accretion rate before the fallback of the most tightly bound material and explicitly express the time-dependent accretion rates before and after $t_{\rm fb}$ as
\begin{linenomath*}
\begin{equation}
    \dot M_{\rm BH} = \frac{\eta_{\rm acc}M_\star}{\mathcal C t_{\rm fb}}\times\begin{cases}
       \left(\frac{t}{t_{\rm fb}}\right)^{-\alpha}, & t<t_{\rm fb}\\
        \left(\frac{t}{t_{\rm fb}}\right)^{-5/3},& t>t_{\rm fb},
    \end{cases}
    \label{eq:accretion_rate}
\end{equation}
\end{linenomath*}
where $0\leq\alpha<1$ is the free early-time accretion index\footnote{\cite{2014ApJ...784...87S} pointed out that a slow-decaying accretion rate is possible due to disk internal kinematic viscosity, depending on the type of polytrope stars.}, the accretion efficiency $\eta_{\rm acc}$ represents the fraction of bounded materials that eventually ends up being accreted to the SMBH, and $\mathcal C \equiv3+2/(1-\alpha)$ is introduced to normalize the total accreted mass, e.g., $\int \dot M_{\rm BH}dt=\eta_{\rm acc}M_\star/2$. The accretion efficiency, $\eta_{\rm acc}$, typically depends on the dynamics of mass fallback and disk formation, since typically a fraction of mass would fallback and end up forming a disk. Nevertheless, for simplicity, we opt for a constant value of $\eta_{\rm acc} = 0.1$~\citep[e.g.,][]{2020ApJ...902..108M}, noting its degeneracy with other parameters. Hence, the accretion rate at $t_{\rm fb}$ can be explicitly written as 
\begin{linenomath*}
\begin{equation}\begin{split}  
    \dot M_{\rm BH}(t_{\rm fb})c^2\simeq &5.5\times10^{46}{~\rm erg~s^{-1}}~\eta_{\rm acc,-1}\mathcal C_{0.7}\\
    &\times f_{T,-1.2}^{-1/2}M_{\rm BH,7}^{-1/2}M_{\star,0.7}^{1.1},
    \end{split}
    \label{eq:acc_fb}
\end{equation}
\end{linenomath*}
which implies the accretion rate is initially in the super-Eddington regime, e.g., $\dot M_{\rm BH}c^2\gtrsim L_{\rm Edd}/\eta_{\rm rad}\simeq1.26\times10^{46}M_{\rm BH,7}\eta_{\rm rad,-1}^{-1}~\rm erg~s^{-1}$, where $L_{\rm Edd}$ is the Eddington luminosity and $\eta_{\rm rad}\sim 0.1$ is the radiation efficiency.

\section{Jet Dynamics}\label{sec:jet_dyn}

In this section, we describe the physical framework of our structured jet model and derive the time evolution of jet Lorentz factors incorporating the continuous energy and mass injections. These derivations will be used in Sec. \ref{sec:results} to compute the time-dependent electromagnetic emissions.

\subsection{The physical picture}

Multiwavelength follow-ups of AT 2022cmc demonstrate distinct signatures between radio, optical and X-ray emissions, which imply that they may originate from different radiation zones. 

Firstly, regarding the spectral energy distributions (SEDs), we find that the radio spectra align with synchrotron self-absorption tails in the electron slow-cooling regime. The optical spectra exhibit good agreement with black body distributions whereas the X-ray spectra are consistent with either a single power-law or a broken power-law distribution predicted by synchrotron radiation \citep{2023NatAs...7...88P, 2024ApJ...965...39Y}. In this case, these emissions likely stem from different physical environments characterized by distinct compactness and magnetic fields. For example, the radio and X-ray emissions may be generated by nonthermal electrons accelerated in extended shocks, while the thermal optical emissions probably originate from a thermal envelope within the accretion disk or a hot corona. 

Secondly, concerning the temporal evolution, the radio signals display a long-lasting lightcurve, contrasting with the rapidly decaying X-ray lightcurves, which implies that these emissions might be produced in different regions governed by disparate kinetic equations and initial conditions. 
A relativistic jet with a Lorentz factor greater than 10 is suggested to explain the bright and hard X-ray emissions. \cite{2024ApJ...963...66Z} also demonstrated that a single wide/slow jet similar to GRB afterglow models is insufficient to reproduce the radio and X-ray observations simultaneously.

\begin{figure}\centering
    \includegraphics[width=0.46\textwidth]{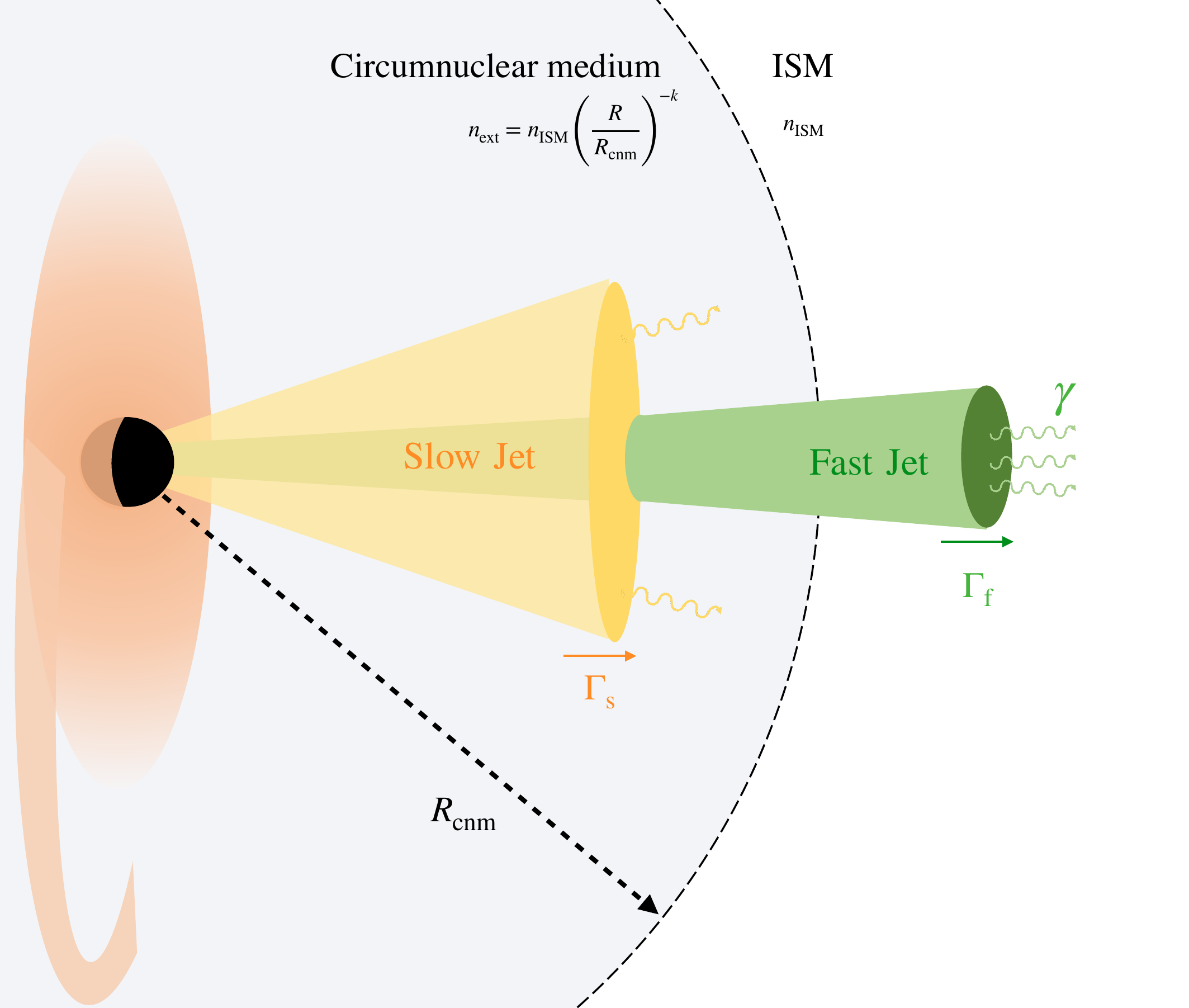}
    \caption{The schematic picture of the structured jet model. A slow jet propagating with Lorentz factor $\Gamma_{\rm s}$ in the circumnuclear material edge ($R_{\rm s} <R_{\rm cnm}$) and a fast jet with Lorentz factor $\Gamma_{\rm f}$ propagating in an ISM ($R_{\rm f}>R_{\rm cnm}$) are illustrated.
    }
    \label{fig:jet_schematic}
\end{figure}

Motivated by these considerations, we consider a time-dependent structured jet model, where the fast/narrow jet and the slow/wide jet are respectively adopted to explain the X-ray and radio lightcurves and spectra at various epochs. Fig. \ref{fig:jet_schematic} illustrates the configuration of our multizone model, depicting an accretion disk, a fast jet with Lorentz factor $\Gamma_{\rm f}$, and a slow jet with Lorentz factor $\Gamma_{\rm s}$. As for the density profile of the external medium, we connect a circumnuclear material edge within the radius $R_{\rm cnm}$ to the interstellar medium. We explicitly write down the density profile (in the units of $\rm cm^{-3}$) in terms of the distance ($R$) to the SMBH,
\begin{linenomath*}
\begin{equation}
    n_{\rm ext}(R)=
    \begin{cases}      
    n_{\rm ISM}\left(\frac{R}{R_{{\rm cnm}}}\right)^{-k}, & R<R_{{\rm cnm}}\\
    n_{\rm ISM}, & R>R_{{\rm cnm}}
    \end{cases}
    \label{eq:n_ext}
\end{equation}
\end{linenomath*}
where $n_{\rm ISM}$ is the number density of ISM, $1.5\leq k\leq2$ is index of the density profile within the material edge radius $R_{\rm cnm}$ as suggested by the radio data fitting \citep[e.g.,][]{2023MNRAS.522.4028M,2024ApJ...965...39Y,2024ApJ...963...66Z}. One potential source of the circumnuclear material is the wind emanating from pre-existing disks. In this scenario, the circumnuclear material radius could extend to $R_{\rm cnm}\sim10^{18}~\rm cm$ before merging with the ISM \citep{2020PhRvD.102h3013Y,2021ApJ...911L..15Y}. In the subsequent subsections, we adopt $k=1.8$ and $R_{\rm cnm}=10^{18}$ cm as fiducial parameters and model the dynamics of jets using $n_{\rm ext}$ defined in Eq. \ref{eq:n_ext}. We will demonstrate in the following subsections that within the data-fitting time window, the fast and slow jets propagate, respectively, in the ISM (e.g., $R_{\rm f}>R_{\rm cnm}$) and the circumnuclear material (e.g., $R_{\rm s}<R_{\rm cnm}$).

Before delving into the kinetic equations for the jets, let us first parameterize the power converted to the fast and slow outflows from accretion using the energy conversion efficiency $\eta_{\rm f}$ and $\eta_{\rm s}$, 
\begin{linenomath*}
\begin{equation}
L_{\rm f/s}=\eta_{\rm f/s}\dot M_{\rm BH}c^2,
\label{eq:L_j}
\end{equation}
\end{linenomath*}
where $\eta_{\rm f/s}\sim\mathcal O(0.1-1)$ represents the fraction of the accreted power is converted to jet.

\subsection{Jet evolution with continuous energy injection}\label{subsec:jet}

We follow the methodology for blastwave dynamics, as in the diagram of GRB afterglow modeling \citep{2013MNRAS.433.2107N,zhang2018physics,2023arXiv231113671Z}, to derive the differential equations governing the evolution of the jet, incorporating time-dependent energy and mass injections. For the fast jet initial Lorentz factor $\Gamma_{\rm f0}\sim\mathcal O(10)$ to be used in the multiwavelength modeling, we estimate the early-time fast jet radius in the order of 
\begin{linenomath*}
\begin{equation}
    R_{\rm f}\sim\frac{2\Gamma_{\rm f0}^2cT_{\rm obs}}{1+z}\simeq2.1\times10^{18}~{\rm cm~}\left(\frac{\Gamma_{\rm f0}}{30}\right)^2\left(\frac{T_{\rm obs}}{1~\rm d}\right),
\end{equation}
\end{linenomath*}
which is larger than $R_{\rm cnm}$ and implies the fast jet is propagating inside the ISM in the time window for lightcurve and SED fitting, e.g., $T_{\rm obs}\gtrsim 10 ~\rm d$. We will show later in this section that the slow jet with initial Lorentz factor $\Gamma_{\rm s0}\sim5$ will decelerate inside the circumnuclear material for $T_{\rm obs}\lesssim100$ d. For both jets, we ignore the adiabatic cooling of the jet as it becomes important only when the density profile index $k\ge3$ is used \cite[see][for details]{2013MNRAS.433.2107N,zhang2018physics}.

Let's consider a general case where the jet of initial Lorentz factor $\Gamma_{0}$ penetrates deeply into the ambient gaseous medium. The jet sweeps up material, resulting in the formation of a forward shock (FS) that accelerates the upstream ISM to the Lorentz factor $\Gamma\lesssim\Gamma_{0}$, and a reverse shock that decelerates the unshocked ejecta from an initial Lorentz factor $\Gamma_{0}$ to $\Gamma$. Fig. \ref{fig:shock_schematic} schematically shows the geometry of the forward shock, reverse shock and the discontinuity between their downstreams. We consider a simplified case that neglects the impact of the reverse shock (RS) on jet evolution. A comprehensive treatment incorporating the reverse shock and a justification for this simplification is provided in Appendix \ref{sec:app_jet}. 

We utilize the fast jet as an exemplar to derive the differential equations describing the jet's evolution. Adopting the approach used in external shock models for GRB afterglow modeling \citep[e.g.,][]{1999MNRAS.309..513H,2012ApJ...752L...8P,2013MNRAS.433.2107N} and neglecting the radiative cooling, we express the total isotropic-equivalent energy of the relativistic jet as 
\begin{linenomath*}
\begin{equation}
    \mathcal E_{\rm f,\rm iso}=\Gamma_{\rm f}M_{\rm ej}c^2+\Gamma_{\rm f}m_{\rm ext}c^2 + \Gamma_{\rm f,\rm eff}\mathcal E_{\rm f,\rm int}',
    \label{eq:E_j}
\end{equation}
\end{linenomath*}
where $M_{\rm ej}$ represents the isotropic-equivalent ejecta mass, $m_{\rm ext}$ is the external mass swept by the outflow, $\mathcal E_{\rm f,\rm int}'=(\Gamma_{\rm f}-1)m_{\rm ext}c^2$ represents the internal energy of the shocked material (downstream) in the jet comoving frame, $\Gamma_{\rm f,\rm eff}=(\hat\gamma\Gamma_{\rm f}^2-\hat\gamma+1)/\Gamma_{\rm f}$ is the effective Lorentz factor and $\hat\gamma=(4+\Gamma_{\rm f}^{-1})/3$ is the adiabatic index taking into account the transition from relativistic to mild-relativistic. Noting the continuous energy and mass injections and the propagation of the outflow, we explicitly write down 
\begin{linenomath*}
\begin{equation}
    M_{\rm ej}=\int dt \frac{L_{\rm f,\rm iso}}{\Gamma_{\rm f0}c^2},~m_{\rm ext}=\int  4\pi R_{\rm f}^2m_pn_{\rm ext}dR_{\rm f},
    \label{eq:M_j}
\end{equation}
\end{linenomath*}
where $L_{\rm f,\rm iso}=L_{\rm f}/(\theta_{\rm f}^2/2)$ is the isotropic jet luminosity given the jet opening angle $\theta_{\rm f}$, and $\Gamma_{\rm f0}$ is the jet initial Lorentz factor before deceleration.
From the perspective of energy conservation, the change of $\mathcal E_{\rm f,\rm iso}$ is
\begin{linenomath*}
\begin{equation}
    d\mathcal E_{\rm f,\rm iso}=c^2dm_{\rm ext} + L_{\rm f,\rm iso}dt.
    \label{eq:d_Ej}
\end{equation}
\end{linenomath*}
In this expression, the first term accounts for the energy by accumulating external mass into the jet whereas the second component demonstrates the persistent energy injection from the central engine. 

\begin{figure}
    \centering
    \includegraphics[width=0.49\textwidth]{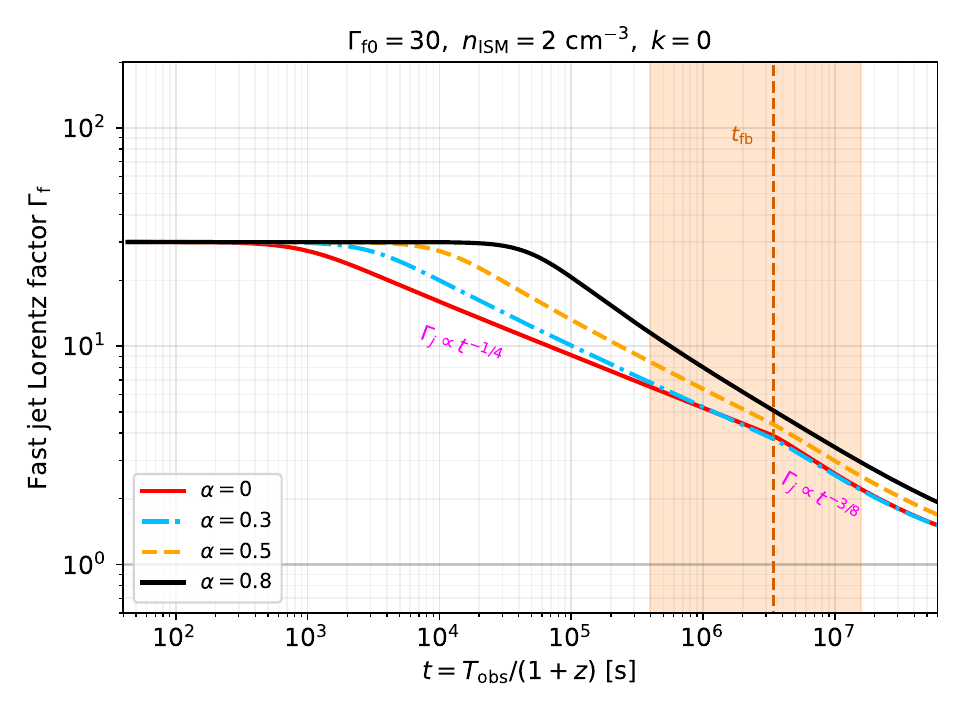}
    \caption{Test time evolution of fast jet Lorentz factor with persistent power and mass injections for the simple case $k=0$. From the thin red curve to the thick black curve, $\alpha$ varies from 0 to 0.8. The vertical orange dashed line shows the fallback time, whereas the orange area represents the time window ($10~{\rm d}\lesssim T_{\rm obs}\lesssim400$ d) for X-ray and radio data fitting.}
    \label{fig:LorentzJ}
\end{figure}

Combining Eqs. \ref{eq:E_j}, \ref{eq:M_j} and \ref{eq:d_Ej}, we obtain the differential equation for the jet deceleration
\begin{linenomath*}
\begin{equation}
    \frac{d\Gamma_{\rm f}}{dm_{\rm ext}}=-\frac{4\Gamma_{\rm f}^5-(5+3A_{\rm inj})\Gamma_{\rm f}^3+\Gamma_{\rm f}}{3M_{\rm ej}\Gamma_{\rm f}^3-2m_{\rm ext}+8\Gamma_{\rm f}^4m_{\rm ext}},
    \label{eq:jet_diff}
\end{equation}
\end{linenomath*}
which resembles the formulation for GRBs \citep[e.g.,][]{2021ApJ...920...55Z}, with the exception of an additional correction factor arising from continuous energy and mass injection,
\begin{linenomath*}
\begin{equation}
    A_{\rm inj}=\left(1-\frac{\Gamma_{\rm f}}{\Gamma_{\rm f0}}\right)\frac{L_{\rm f,\rm iso}}{c^2}\left(\frac{dm_{\rm ext}}{dt}\right)^{-1}.
\end{equation}
\end{linenomath*}
To numerically solve Eq. \ref{eq:jet_diff}, we connect the jet radius to the SMBH-rest frame time $t$, e.g.,
\begin{linenomath*}
\begin{equation}
    dR_{\rm f} = \frac{\beta_{\rm f}cdt}{1-\beta_{\rm f}},
    \label{eq:R_j}
\end{equation}
\end{linenomath*}
where $\beta_{\rm f}=\sqrt{1-\Gamma_{\rm f}^{-2}}$. 

To test the impact of time-dependent power injection rates, we show in Fig. \ref{fig:LorentzJ} the time evolution of jet Lorentz factor using the simple density profile with $k=0$ and $n_{\rm ISM}=2~{\rm cm^{-3}}$ for the fast jet case. We fix the initial jet Lorentz factor to be $\Gamma_{\rm f0}=30$ and use the values $\eta_{\rm f}=0.12$ and $\theta_{\rm f}=0.15$ obtained from the X-ray data fitting in Sec. \ref{sec:results}. From thin red curve to the thick black curve, the parameter $\alpha$ varies from 0 to 0.8. The vertical orange dashed line depicts the fallback time whereas the yellow area represents the time window of the X-ray and radio data to be fitted. Initially, the external mass $m_{\rm ext}$ is not high enough to decelerate the jet and the Lorentz factor remains a constant level, known as the `coasting phase'. The deceleration occurs at $t_{\rm dec}$ when $M_{\rm ej}/\Gamma_{\rm f0}\simeq m_{\rm ext}$ is satisfied. The decaying slope of $\Gamma_{\rm f}$ before $t_{\rm fb}$ for each curve in Fig. \ref{fig:LorentzJ} aligns well with the theoretical estimation $\Gamma_{\rm f}\propto t^{-(2+\alpha)/8}$ derived from $\int L_{\rm f,\rm iso}dt\sim4\pi R_{\rm f}^3\Gamma_{\rm f}^2m_pc^2/3$. In this regime, we find the correction factor to be $A_{\rm inj} \lesssim 0.1$. The increasing $M_{\rm ej}$ primarily leads to the modifications from the continuous injection.

For $t>t_{\rm fb}$, the energy injection rate decays much faster and the $\Gamma_{\rm f}$ enters the Blandford-McKee (BM) self-similar regime \citep{1976PhFl...19.1130B}, e.g., the $\Gamma_{\rm f}\propto R_{\rm f}^{-2/3}\propto t^{-3/8}$, predicted by a fixed $\mathcal E_{\rm f,\rm iso}\sim 4\pi R_{\rm f}^3\Gamma_{\rm f}^2n_{\rm ext}m_pc^2/3$. In this picture, we typically expect two break points at $t_{\rm dec}$ and $t_{\rm fb}$, similar to the red curve in Fig. \ref{fig:LorentzJ}. However, for a larger $\alpha$, $t_{\rm dec}$ becomes closer to $t_{\rm fb}$ and the decaying slope $(2+\alpha)/8$ approaches to $3/8$, which as a result makes the second break at $t_{\rm fb}$ less prominent. We also observe that $\Gamma_{\rm f}(t)$ is very sensitive to $\alpha$ as it approaches to 1.

The derivation presented above is also applicable to slow jets by substituting the corresponding physical quantities with those for slow jets., e.g., $\Gamma_{\rm s}, L_{\rm s}$ and $\theta_{\rm s}$. For a slow jet with $\Gamma_{\rm s0}=4$, we estimate the deceleration time $t_{\rm dec,s}\sim 17$ d using
\begin{linenomath*}
\begin{equation}
    \int L_{\rm s,\rm iso}dt = \frac{4\pi}{3-k}\Gamma_{\rm s0}^2m_pn_{\rm ext}(R_{\rm s})R_{\rm s}^3c^2
\end{equation}
\end{linenomath*}
and the radius $R_{\rm s}\approx2\Gamma_{\rm s0}^2ct$. We can also estimate the slow jet radius at $T_{\rm obs}=100$ d to be $R_{\rm s}\lesssim R_{\rm cnm}$ which implies that the slow jet is propagating within the circumnuclear medium characterized by the index $k=1.8$. 

{We examined the time evolution of the slow jet and found that it becomes mildly relativistic, e.g., $\Gamma_{\rm s} \simeq 1.4$, by the end of the data fitting time window.  This occurs because the CNM density profile $n_{\rm ext} \propto R^{-1.8}$ results in a slowly decelerating jet. In this case, our treatments for relativistic jets still apply to the slow jet.}

So far, we have derived the differential equations for TDE jet evolution, taking in to account the continuous injection rate defined in Eq. \ref{eq:accretion_rate} and Eq. \ref{eq:L_j}. In Sec. \ref{sec:results}, we will apply this model to both fast and slow jets with different initial conditions and jet parameters. Subsequently, we will compute the synchrotron and inverse Compton emissions in their respective forward shock and reverse shock regions.

\section{Multizone and Multiwavelength modeling}\label{sec:results}

In this section, we model the emissions from the forward and reverse shocks for both fast and slow jets. We apply the structured jet model to fit the X-ray (0.3-10 keV) and radio/millimeter (15.5 GHz and 225 GHz) lightcurves and spectra. The jets considered in this work are all on-axis. We use two sets of jet parameters ($\eta_{\rm f}$, $\theta_{\rm f}$, $\Gamma_{\rm f0}$) and ($\eta_{\rm s}$, $\theta_{\rm s}$, $\Gamma_{\rm s0}$) to describe the time evolutions of the fast and slow jets, as summarized in Table \ref{tab:params}.

Regarding the thermal optical emissions, which may stem from {the thermalization of emission} from an accretion disk or a hot corona, we consider them as upper limits within the multizone model.

\subsection{Jet forward shock and reverse shock modeling}\label{subsec:FS_RS_model}

For simplicity, we continue using the fast jet as an example, noting that the forward and reverse shock models presented here are also applicable to slow jets by substituting the corresponding physical quantities with those for slow jets.

{We also note that we use the single-zone approximation for each shock component, in the sense that we do not consider integration over the equivalent arrival time surface. Our model is regarded as a mulitizone model in light of forward and reverse shock regions of fast and slow jets.}

{\bf Forward shock (FS) -- }In the jet forward shock model, given the external particle number density $n_{\rm ext}$ and $\Gamma_{\rm f}$, we parameterize the magnetic field strength of the downstream  magnetic field as $B_{\rm f,\rm fs}=[{32\pi\Gamma_{\rm f}(\Gamma_{\rm f}-1)\epsilon_{B}^{\rm fs}n_{\rm ISM}m_pc^2}]^{1/2}$, where $\epsilon_{B}^{\rm fs}$ represents the fraction of internal energy density that goes into the magnetic field energy density. We consider the shock-accelerated nonthermal electrons, described by a power-law injection rate, e.g., $\dot Q_{e,\rm fs}(\gamma_e)\propto \gamma_e^{-s}$, where $\gamma_e$ is the electron Lorentz factor and $s\ge2.0$ is the spectral index. To normalize the injection rate, we introduce the number fraction ($f_{e}^{\rm fs}$) of the downstream electrons that are accelerated and the energy fraction ($\epsilon_{e}^{\rm fs}$) of the internal energy that are deposited to nonthermal electrons. In this approach, we infer the minimum Lorentz factor for injected electrons, 
\begin{linenomath*}
\begin{equation}
    \gamma_{e,m}^{\rm fs}=(\Gamma_{\rm f}-1)g(s)\frac{\epsilon_{e}^{\rm fs}}{f_{e}^{\rm fs}}\frac{m_p}{m_e},
    \label{eq:gamma_e_m_fs}
\end{equation}
\end{linenomath*}
where $g(s)=(s-2)/(s-1)$ for $s>2.0$ and $g(s)\sim\mathcal O(0.1)$ for $s=2.0$. We then normalize $\do Q_{e,\rm fs}$ via 
\begin{linenomath*}
\begin{equation}
    (4\pi R_{\rm f}^2t_{\rm f,{\rm dyn}}'c)\int \dot Q_{e,\rm fs} d\gamma_e=\frac{4\pi f_{e}^{\rm fs}R_{\rm f}^3n_{\rm ext}}{3t_{\rm f,\rm dyn}'},
    \label{eq:ele_inj_norm_fs}
\end{equation}
\end{linenomath*}
where $t_{\rm f,\rm dyn}'=R_{\rm f}/(\Gamma_{\rm f}c)$ is the dynamic time measured in the comoving frame.

\begin{deluxetable}{ccchlDlc}
\tablenum{1}
\tablecaption{Physical (fiducial and fitting) parameters for the structured jet model \label{tab:params}}
\tablewidth{0pt}
\tablehead{
\colhead{\bf Category} &\colhead{\bf Parameter} & \colhead{\bf Value} & \nocolhead{Common} & 
}
\startdata
 {}  & $M_{\rm BH}$ & $10^7~M_\odot$ \\
 {} & $M_\star$ & $5~M_\odot$\\
 {Fiducial} & $\eta_{\rm acc}$ & 0.1\\
 {} & {$R_{{\rm cnm}}$} & $10^{18}~{\rm cm}$\\
 {} & $k$ & 1.8\\
 \hline
 {\bf Fitting parameters}\\
 \specialrule{.1em}{.06em}{.06em} 
 {} &$\alpha$ & 0.8  \\
 {Universal} & $n_{\rm ISM}$ & 2.0$~\rm cm^{-3}$\\
  {} & $s$ & 2.3\\
 \hline
 {} & {$\eta_{\rm f,s}$} & {0.12, 0.04}\\
 {Fast, slow jets} &{$\theta_{\rm f,s}$} & {0.15, 0.3}\\
 {} & {$\Gamma_{\rm f0, s0}$} & {30, 4.0}\\
 \hline
 {} & $\epsilon_{e}^{\rm fs, rs}$ & 0.1, 0.2 \\
 {FS, RS} & $\epsilon_{B}^{\rm fs, rs}$ & $3.0\times10^{-3}$, 0.1\\
 {} & $f_{e}^{\rm fs,rs}$ & {1.0, $1.5\times10^{-3}$}\\
 \hline
\enddata
\end{deluxetable}

{\bf Reverse shock (RS) -- }For the reverse shock, we parameterize the magnetic field strength and electron injection rate for the reverse shocks, e.g., $B_{\rm f, rs}$ and $\dot Q_{e,\rm rs}$. In contrast to the forward shock scheme, the relative Lorentz factor between reverse shock up and downstreams is $\Gamma_{\rm f,\rm rs-rel}\approx (\Gamma_{\rm f0}/\Gamma_{\rm f}+\Gamma_{\rm f}/\Gamma_{\rm f0})/2$. Similar to the forward shock case, we define the reverse shock parameters $\epsilon_{e}^{\rm rs}$, $f_{e}^{\rm rs}$, and $\epsilon_{B}^{\rm rs}$. In this case, the magnetic field strength of the reverse shock region can be written as 
\begin{linenomath*}
\begin{equation}
B_{\rm f,\rm rs}=\sqrt{32\pi\epsilon_{B}^{\rm rs}\Gamma_{\rm f,\rm rs-rel}(\Gamma_{\rm f,\rm rs-rel}-1)n_{\rm f0}'m_pc^2},
\end{equation}
\end{linenomath*}
where $n_{\rm f0}'=L_{\rm f,\rm iso}/(4\pi \Gamma_{\rm f0}^2R_{\rm f}^2m_pc^3)$ is the comoving upstream number density of the reverse shock.
The minimum Lorentz factor of injected electrons in reverse shocks, e.g., $\gamma_{e,m}^{\rm rs}$ can be obtained by replacing $\Gamma_{\rm f}$ with $\Gamma_{\rm f,\rm rs-rel}$ in Eq. \ref{eq:gamma_e_m_fs}. Moreover, the particle number injection rate $\dot N_{\rm rs}=L_{\rm f,\rm iso}/(\Gamma_{\rm f0}m_pc^2)$ should be used to normalize $\dot Q_{e,\rm rs}$, which can be expressed explicitly as
\begin{linenomath*}
\begin{equation}  
(4\pi R_{\rm f}^2t_{\rm f,\rm dyn}'c)\int \dot Q_{e,\rm rs}d\gamma_e = f_{e}^{\rm rs}\dot N_{\rm rs}=\frac{f_{e}^{\rm rs}L_{\rm f,\rm iso}}{\Gamma_{\rm f0}m_pc^2}.
\label{eq:ele_inj_norm_rs}
\end{equation}
\end{linenomath*}

\begin{figure*}[htp]
    \centering
    \includegraphics[width=0.99\textwidth]{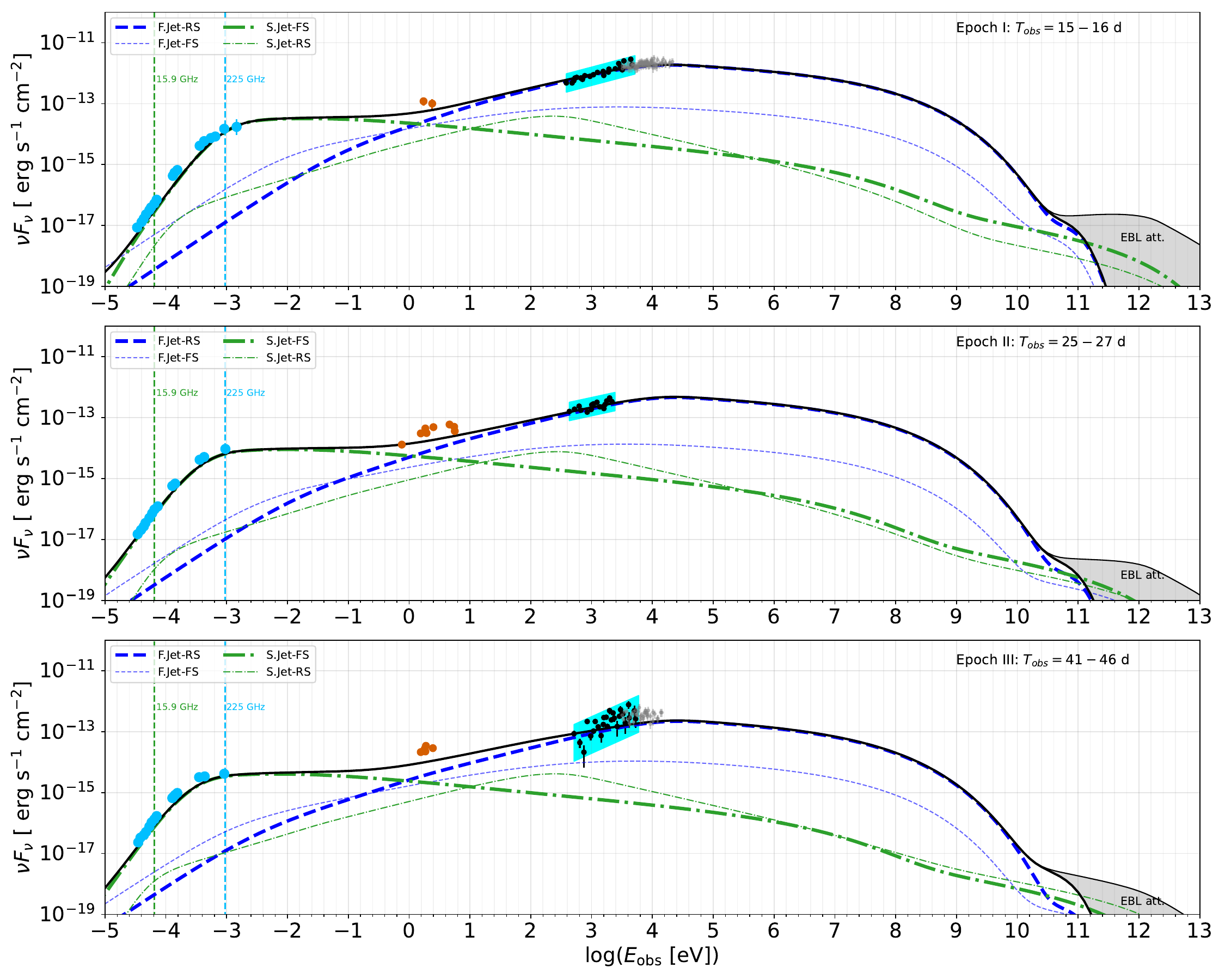}
    \caption{Spectral fitting for three epochs. The `SY/IC' spectra of the fast/slow jet forward (F.Jet/S.Jet-FS) and reverse shocks (F.Jet/S.Jet-RS), are illustrated. The orange points depict the energy fluxes in optical bands, which are considered as the upper limits. The thin and thick solid black curves correspond to the overall SED before and after applying $\gamma\gamma$ attenuation with EBL. The radio/millimeter and X-ray data are depicted as blue and black dots, respectively \citep{2022Natur.612..430A,2023NatAs...7...88P}. Additionally, the hard X-ray energy fluxes \citep{2024ApJ...965...39Y} are also shown as gray points.}
    \label{fig:specs}
\end{figure*}

Given the injected electron rates and the magnetic fields in forward and reverse shocks, we use the AM$^3$ software \citep{2023arXiv231213371K} to model the synchrotron and synchrotron self-Compton emissions in these two regions by numerically solving the corresponding time-dependent transport equations for electrons in the comoving frame,
\begin{linenomath*}
\begin{equation}
    \frac{\partial n_e'}{\partial t'}=\dot Q_{e}-\frac{\partial}{\partial \gamma_e}\left(\dot \gamma_en_e'\right)-\frac{n_e'}{t_{\rm f,\rm dyn}'}.
\end{equation}
\end{linenomath*}
In this equation, $n_e'$ and $t'$ represents the electron number density (differential in Lorentz factor and volume) and time measured in jet comoving frame, $\dot\gamma_e\sim\gamma_e(t_{e,\rm sy}'^{-1}+t_{e,\rm ic}'^{-1})$ is the electron energy loss rate due to synchrotron and inverse Compton radiation in the synchrotron self-Compton diagram. In addition, we self-consistently estimate the electron maximum Lorentz factor by balancing the efficient acceleration rate $t_{e,\rm acc}'^{-1}\sim eB_{\rm f}/(\gamma_em_ec)$ and the cooling rate $t_{e,c}'^{-1}=t_{e,\rm sy}'^{-1}+t_{e,\rm ic}'^{-1}$. To get the observed photon spectra, we convert the obtained comoving photon density spectrum $n_\gamma'=d^2N_\gamma/(d\ln E_\gamma'dV')$ in the units of $\rm cm^{-3}$ to the flux in the observer's frame via
\begin{linenomath*}
\begin{equation}
    \nu F_\nu(E_{\gamma,\rm obs})=f_{\rm br}\Gamma_{\rm f}^2\left(\frac{R_{\rm f}^2}{d_L^2}\right)cE_{\gamma}'n_{\gamma}'\exp(-\tau_{\rm EBL}),
\end{equation}
\end{linenomath*}
where the observed and comoving energies are connected by $E_{\gamma,\rm obs}=\Gamma_{\rm f}E_{\gamma}'/(1+z)$, $f_{\rm br}=1/[1+(\Gamma_{\rm f}\theta_{\rm f})^{-2}]$ accounts for the jet break correction, and we also applied the correction of $\gamma\gamma$ absorption attributed to the attenuation with extragalactic background light (EBL) during propagating from $z=1.19~(d_L\simeq8.4$ Gpc) to the Earth.

\subsection{Fast jet: X-ray data fitting}
\label{subsec:X_fitting}

We apply the dynamics of the fast and slow jet (Sec. \ref{subsec:jet}) together with the modeling of the forward and reverse shocks to explain the measured 0.3-10 keV lightcurve and the X-ray spectra in multiple epochs \citep{2022Natur.612..430A,2023NatAs...7...88P, 2024ApJ...965...39Y}, e.g., $T_{\rm obs} = 15-16~{\rm d},~25-27~{\rm d},~41-46~{\rm d}$. To reduce the free parameters, we assume the forward shocks of the fast and slow jets share the same $\epsilon_{e}^{\rm fs}$, $\epsilon_{B}^{\rm fs}$ and $f_{e}^{\rm fs}$, whereas all reverse shocks have the same $\epsilon_{e}^{\rm rs}$, $\epsilon_{B}^{\rm rs}$ and $f_{e}^{\rm rs}$.

We fix the spectral index of injected electrons to be $s=2.3$ during our calculation. The fiducial parameters for the TDE accretions and the external density profile, together with the physical parameters of the jets and the reverse/forward shocks obtained by fitting the X-ray and radio/millimeter spectra and lightcurve, are presented in Table \ref{tab:params}.

\begin{figure}[t]
    \centering
    
    \includegraphics[width=0.49\textwidth]{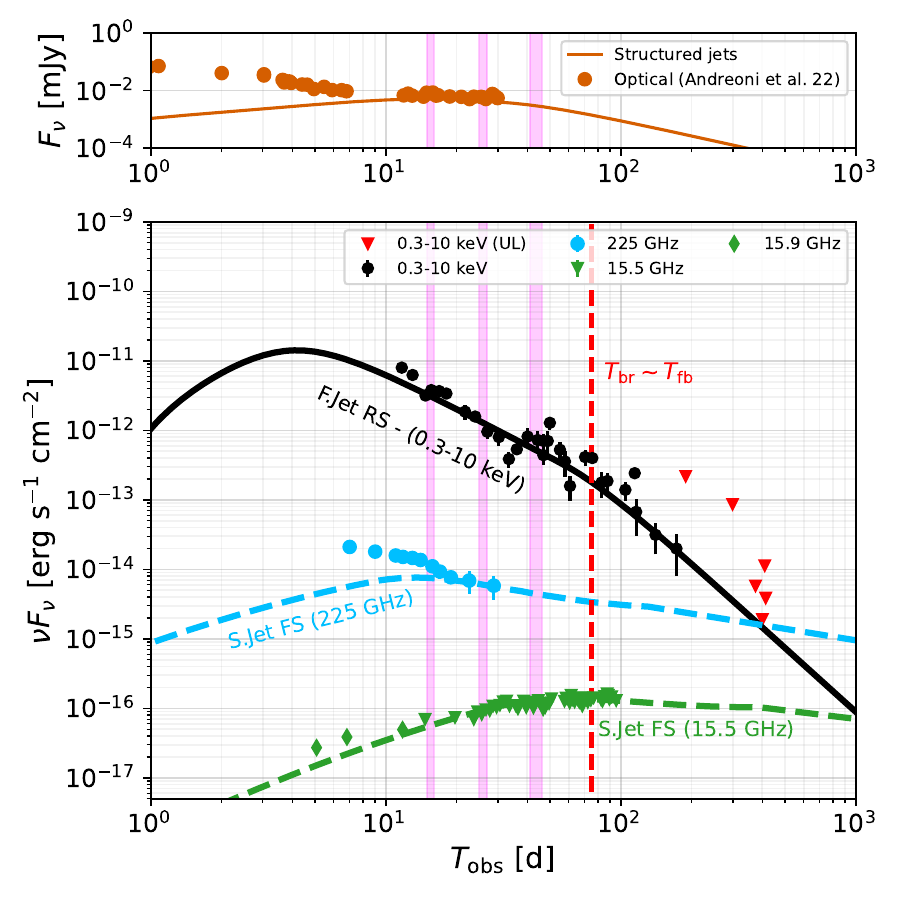}

    \caption{{\emph{Top panel:} Comparison of the model-predicted optical lightcurve (orange solid curve) to the observation (orange points).} \emph{Bottom panel:} fitting of nonthermal X-ray and radio lightcurves using the parameters in Table \ref{tab:params}. 
    The solid black line represents the 0.3-10 keV X-ray afterglow from the fast jet's reverse shock model (F.Jet RS). The blue and green dashed curves depict the 225 GHz and 15.9 GHz lightcurves obtained from the slow jet forward shock (S.Jet FS) scenario. The vertical magenta areas represent the three epochs for SED fitting in Fig. \ref{fig:specs}. The latest X-ray upper limits (ULs) are depicted as red triangles. The vertical red line represents the fast jet break time ($T_{\rm br}$), which is close to the fallback time ($T_{\rm fb}$). Data sources: \cite{2022Natur.612..430A,2023NatAs...7...88P, 2024ApJ...965...39Y, 2024arXiv240410036E}}
    \label{fig:lightcurve}
\end{figure}

Fig. \ref{fig:specs} shows the fitting to the SED in three epoches, $T_{\rm obs}=12-15{~\rm d},~25-27~{\rm d},~41-46~{\rm d}$, respectively in the top, middle and bottom panels. The X-ray (black points), optical (orange points), and radio/millimeter (blue points) are sourced from \cite{2022Natur.612..430A,2023NatAs...7...88P}, whereas the hard X-ray data \citep[gray points,][]{2024ApJ...965...39Y} at the closest observation windows are also shown. In the fast jet (denoted as `F.Jet') scenario, the thin and thick blue dashed curves depict the combined synchrotron and inverse Compton (denoted as `SY/IC') emissions from the forward shock (FS) and reverse shock (RS) regions. From this figure, we find that, by adopting the parameters in Table \ref{tab:params}, the reverse shock fast-cooling synchrotron spectra are consistent with the observed X-ray data. We also observe that the forward shock is subdominant using the parameters for radio data fitting. The forward shock and reverse shock spectra are also consistent with the analytical predictions in Appendix \ref{sec:app_analytical}.

In Fig. \ref{fig:lightcurve}, we present the fitting to the measured X-ray lightcurve \citep{2024ApJ...965...39Y} using the jet scenario in the lower panel. Our results indicate that emissions from the fast jet reverse shock could account for the fast-decaying X-ray lightcurve, since the decaying accretion rate defined in Eq. \ref{eq:accretion_rate} can directly influence the injection rate of the accelerated electrons. As we observed that the forward shock is sub-dominant in the X-ray bands and predicts a more flat lightcurve, its lightcurve is not displayed in this figure. 

To comprehend the temporal evolution of these regions, we derive the analytical X-ray lightcurves considering the time-dependent energy injection luminosity before and after the fallback time. Requiring $\Gamma_{\rm f}=1/\theta_{\rm f}$, we infer the fast jet break time to be $t_{\rm br}\simeq 2.8\times10^6~{\rm s}$ for $\theta_{\rm f}=0.15$, which is close to the fallback time $t_{\rm fb}$. For simplicity, in the following discussions, we do not distinguish $t_{\rm fb}$ and $t_{\rm br}$, and treat the jet as post-break and apply the steepening factor $f_{\rm br}=(\Gamma_{\rm f}\theta_{\rm f})^2\propto T_{\rm obs}^{-3/4}$ to the lightcurve after $t_{\rm fb}$. In this case, the analytical X-ray lightcurves for both forward and reverse shocks in the fast cooling regime can be written respectively as
\begin{linenomath*}
\begin{equation}
    \nu F_\nu^{\rm (fs)}\propto \begin{cases}
        f_{\rm br}T_{\rm obs}^{-\alpha+1-s/2},&T_{\rm obs}<T_{\rm br}\simeq T_{\rm fb}\\
        f_{\rm br}T_{\rm obs}^{-1}, & T_{\rm obs}>T_{\rm br}\simeq T_{\rm fb},
    \end{cases}
\end{equation}
\end{linenomath*}
and 
\begin{linenomath*}
\begin{equation}
    \nu F_\nu^{\rm (rs)}\propto\begin{cases}
        T_{\rm obs}^{-[5\alpha+\alpha(s-1)]/4}, &T_{\rm obs}<T_{\rm br}\simeq T_{\rm fb}\\
        T_{\rm obs}^{-(2s+25)/12}, &T_{\rm obs}>T_{\rm br}\simeq T_{\rm fb}.\\
    \end{cases}
    \label{eq:rs_lightcurves}
\end{equation}
\end{linenomath*}
where $T_{\rm br}=(1+z)t_{\rm br}$, $T_{\rm fb}=(1+z)t_{\rm fb}$, and a detailed derivation is provided in Appendix \ref{sec:app_analytical}. They are consistent with the GRB afterglow lightcurves for both forward shocks and reverse shocks \cite[see][for details]{zhang2018physics}. Since we adopted the approximations, $\Gamma_{\rm f, rs-rel}\sim\Gamma_{\rm f, rs-rel}-1\sim \Gamma_{\rm f0}/(2\Gamma_{\rm f})$, to obtain Eq. \ref{eq:rs_lightcurves}, the numerical time decay slope could be steeper by a correction factor $\sim T_{\rm obs}^{-0.2}-T_{\rm obs}^{-0.4}$. The analytical lightcurves closely match the numerical results obtained using $\alpha=0.8$ in Fig. \ref{fig:lightcurve}. Specifically, the jet break time in the observer's frame, $T_{\rm br}\simeq 72~\rm d$, and the steepened temporal evolution for $s=2.3$, such as $\propto T_{\rm obs}^{-2.7}-T_{\rm obs}^{-2.9}$, after $T_{\rm br}$, are consistent with the X-ray fluxes in the time window $T_{\rm br}<T_{\rm obs}<200~\rm d$ and the late-time upper limits \citep[illustrated as red triangles,][]{2024arXiv240410036E} extending to $T_{\rm obs}\sim 400$ d. Conversely, the forward shock is disfavored due to its relatively flat temporal evolution and negligible flux in X-ray bands. 

{We emphasize that there is no necessary physical causation for the relation $T_{\rm br}\simeq T_{\rm fb}$. The mass fallback time is determined by the SMBH mass and the structure of the disrupted star, such as $M_\star$ and $R_{\star}$. Since we use fiducial parameters for these quantities, the mass fallback time is roughly fixed and does not depend on our fitting parameters. As for the jet break time, we found that $T_{\rm br} \sim T_{\rm fb}$ is favored for fitting the X-ray lightcurve. For example, a later $T_{\rm br}$ would exceed the late-time X-ray upper limits, whereas an earlier $T_{\rm br}$ would underestimate the X-ray flux at $T_{\rm obs}\sim100$ d.} 

\subsection{Slow jet: radio/millimeter data fitting}
\label{subsec:X_fitting}

Recently, \cite{2023MNRAS.522.4028M}, \cite{2024ApJ...965...39Y}, and \cite{2024ApJ...963...66Z} demonstrated that jet forward shocks, similar to those in GRB afterglow models, can interpret the radio observations of AT 2022cmc. The radio spectra and lightcurves are well described by a decelerated, mildly relativistic jet with an initial Lorentz factor of $\sim3-10$ in a circumnuclear medium characterized by an index of $1.5\lesssim k\lesssim2$. In this work, we aim to provide a more comprehensive model of the multiwavelength emissions from jetted TDEs by considering a slow jet (in addition to the fast jet used for X-ray data fitting) propagating within $R_{\rm cnm}$. This approach is consistent with previous studies. The major difference in our work is the inclusion of persistent energy injection due to long-lasting accretion activities, as discussed in Sec. \ref{subsec:jet}.

Similar to the fast jet scenarios, we self-consistently compute the time-dependent synchrotron and inverse Compton emissions in the forward shock and reverse shock regions, as discussed in Sec. \ref{subsec:FS_RS_model}. By fitting the radio observations, we obtain the slow jet parameters ($\eta_{\rm s}$, $\theta_{\rm s}$, and $\Gamma_{s0}$) and the forward shock parameters ($\epsilon_e^{\rm fs}$, $\epsilon_B^{\rm fs}$ and $f_e^{\rm fs}$), as summarized in Table \ref{tab:params}. Comparing to the fast $\Gamma_{\rm f0}=30$ and narrow ($\theta_{\rm s}=0.15$) jet considered in Sec. \ref{subsec:X_fitting}, a slow ($\Gamma_{\rm s0}=4$) and wide ($\theta_{\rm s}=0.3$) jet is favored to fit the radio data. For completeness, we incorporate the reverse shock parameters obtained from X-ray data fitting in the fast jet scenario to account for the contribution from the slow jet reverse shock regions.

The fitted forward shock SY/IC spectra in the slow jet scenario are shown as the thick green dashed-dotted curves in Fig. \ref{fig:specs} together with the observed radio/millimeter SED (blue points) and the reverse shock componets (thin green dashed-dotted curves). In Fig. \ref{fig:lightcurve}, we also compare our model-predicted 15.5 GHz (green curve) and 225 GHz (blue curve) lightcurves with the corresponding observations. Our results demonstrate that the forward shock of the continuously powered slow jet can reproduce the observed radio/millimeter spectra and lightcurves. 
Our results and parameters are consistent with \cite{2023MNRAS.522.4028M}, and the cumulative slow jet energy
\begin{linenomath*}
\begin{equation}
    \mathcal E_{\rm s,iso}=\frac{2\eta_{\rm s}\dot M_{\rm BH}(t_{\rm fb})c^2t_{\rm fb}}{(1-\alpha)\theta_{\rm s}^2}\sim8\times10^{53}~\rm erg.
\end{equation}
\end{linenomath*}

To understand the radio spectra fitting, we estimate the synchrotron characteristic frequency and the cooling frequency in the forward shock region to be respectively,
\begin{linenomath*}
\begin{equation}
\begin{split}
\nu_m=\frac{3\Gamma_{\rm s}(\gamma_{e,\rm m}^{\rm fs})^2eB_{\rm s,fs}}{4\pi(1+z)m_ec}&\simeq 3.8\times10^2{~\rm GHz}\\
&\times\left(\frac{T_{\rm obs}}{15~\rm d}\right)^{-0.9}\left(\frac{\Gamma_{\rm s}}{3}\right)^{2.2}
\end{split}
\end{equation}
\end{linenomath*}
and
\begin{linenomath*}
\begin{equation}
\begin{split}
\nu_c=\frac{3\Gamma_{\rm s}\gamma_{e,c}^2eB_{\rm s,fs}}{4\pi(1+z)m_ec}&\simeq 3.1\times10^4{~\rm GHz}\\
&\times\left(\frac{T_{\rm obs}}{15~\rm d}\right)^{0.7}\left(\frac{\Gamma_{\rm s}}{3}\right)^{1.4},
\end{split}
\end{equation}
\end{linenomath*}
where $\gamma_{e,c}\sim6\pi m_ec/(\sigma_TB_{\rm s,fs}^2\Gamma_{\rm s}t)$ is the electron cooling Lorentz factor and $\sigma_T$ is the Thomson cross section.
The wind SY/IC spectra in Fig. \ref{fig:specs} are consistent with the weak synchrotron self-absorption (SSA) in the electron slow-cooling regime, e.g., the absorption frequency $\nu_a<\nu_m<\nu_c$, characterized by a steep increasing tail, e.g., $\nu F_\nu\propto \nu^{3}$ for $\nu<\nu_a$. {Another mechanism that may prevent the radio emission to escape is the free-free absorption due to the thermal electrons in the external medium, which typically requires very high $n_{\rm ext}$. Applying Eq. \ref{eq:n_ext}, we find that the free–free optical depth, $\tau_{\rm ff}\propto n_{\rm ext}^2$ \citep[e.g.,][]{2017ApJ...836L...6M}, is extremely low, and conclude that this effect is negligible in radio data fitting.}

The slow jet forward shock model can explain the 225 GHz and 15.5 GHz fluxes and lightcurves very well after $T_{\rm obs}=15$ d. However, the early stage radio/millimeter fluxes for $T_{\rm obs}<15$ d are underestimated by the slow jet forward shock model, as illustrated in Fig. \ref{fig:lightcurve}. This suggests that the early-time radio emissions may originate from different zones, such as structured two-component outflows \citep{2024ApJ...963...66Z,2024arXiv240413326S}. 
Regarding the slow jet reverse shock component, the use of a very small $f_e^{\rm rs}$ results in a very high minimum Lorentz factor for the accelerated electrons. Consequently, the corresponding SY/IC spectra peak at much higher energies compared to the forward shock case, e.g., the thin green dashed-dotted curves in Fig. \ref{fig:specs}, and the reverse shock contribution to the radio/millimeter energy fluxes is minimal.

{\bf Overall SEDs --} We have presented a comprehensive discussion of the forward shock and reverse shock components in the fast and slow jets. Combining the emissions from these zones, we obtain the overall SED for the three epochs before and after applying the EBL absorption. These are respectively represented by the thin and thick black solid curves in Fig. \ref{fig:specs}. {For reference, the thermal optical data (orange points) are also included as the upper limits, since the UV/optical light may come from difference zones, such as a thermal envelope with radius $\sim10^{15}$ cm \citep{2024ApJ...965...39Y}. The contribution of the jets to the optical bands is consistent with optical observations (see also the upper panel of Fig. \ref{fig:lightcurve}), which further supports the fast cooling synchrotron emission as the origin of the X-rays}. Importantly, our multizone model, incorporating the forward and reverse shocks of both fast and slow jets, effectively describes multi-wavelength observations, encompassing radio and X-ray spectra, along with their corresponding lightcurves.

\section{Discussion}\label{sec:discussion}

{We have analyzed both forward shock and reverse shock scenarios for both fast and slow jets, identifying the reverse shock of the fast jet and the forward shock of the slow jet regions as promising X-ray and radio sources for the jetted TDE AT 2022cmc. Their primary advantages are summarized below:
\begin{itemize}
    \item {\bf Fast jet reverse shock}: The combination of continuously decaying energy injection and deceleration in the ISM leads to a rapidly decaying X-ray lightcurve, consistent with observations. {Fitting the spectra requires a low value of $f_{e}^{\rm rs}=1.5\times10^{-3}$, which predicts a high $\gamma_{e,m}^{\rm rs}$. This value is consistent with values suggested from the reverse shock model for early GRB afterglow emission~\citep{Genet:2007nb} and 
    could be related to the injection physics of the reverse shock of magnetically-dominated jets.}
    
    \item{\bf Slow jet forward shock}: The typical values of $\epsilon_B^{\rm fs}=3.0\times10^{-3}$ and $f_e^{\rm fs}=1$, consistent with GRB afterglow modeling, along with the presence of a dense external medium, facilitate the consistency of SSA with the radio spectra. Additionally, the radio/millimeter lightcurves can be effectively modeled by the deceleration of the jet within the circumnuclear medium ($R_{\rm s}\lesssim R_{\rm cnm}$).
\end{itemize}
}

\subsection{Fitting parameters}

{\bf Mass of the disrupted star --} To construct the accretion history of AT 2022cmc, we {have initially assumed} a disrupted star of 5 $M_\odot$ in Eq. \ref{eq:accretion_rate}. A lower limit for $M_\star$ can be estimated based on the isotropic equivalent X-ray luminosity, e.g., $L_{X,\rm iso}\sim3\times10^{47}{\rm erg~s^{-1}}(T_{\rm obs}/5~\rm d)^{-2}$ \citep{2022Natur.612..430A}, which implies an X-ray energy $E_{X,\rm iso}\gtrsim1.3\times10^{53}$ erg. Using the energy conversion efficiencies, the mass of the disrupted star can be estimated as 
\begin{linenomath*}
\begin{equation}
    M_\star\sim \frac{2f_{\rm bol}f_bE_{X,\rm iso}}{\eta_{\rm acc}\eta_{\rm f}\epsilon_{e}^{\rm fs}c^2}\gtrsim3.3M_\odot\eta_{\rm acc,-1}^{-1}\eta_{\rm f,-1}^{-1},
\end{equation}
\end{linenomath*}
where $f_{\rm bol}\sim3-4$ is bolometric correction factor and $f_{b}=\theta_{\rm f}^2/2$ is the jet beaming factor. Additionally, when lower accretion and jet efficiencies  are used, a significantly more massive star is needed to explain the bright X-ray emissions. As a result, the likelihood of such disruptions would decrease notably. Thus, our fiducial values for $M_\star$, $\eta_{\rm acc}$, and $\eta_{\rm f}$ are reasonable rather than optimistic.

{{\bf Number of parameters -} In addition to the fiducial parameters fixed/constrained by observations, e.g., $M_{\rm BH}$ and $M_\star$, and by theoretical estimations, e.g., $R_{\rm cnm},~k$, and $\eta_{\rm acc}$, there are three global parameters: $n_{\rm ISM},~\alpha$ and the spectral index $s$. For the X-ray data fitting with the fast jet, the free parameters are: jet evolution parameters ($\Gamma_{\rm f0}$, $\eta_{\rm f}$, $\theta_{\rm f}$) and reverse shock parameters ($\epsilon_{e}^{\rm rs}$, $\epsilon_{B}^{\rm rs}$, $f_e^{\rm rs}$). Similarly, for the slow jet and radio data fitting, there are six parameters: ($\Gamma_{\rm s0}$, $\eta_{\rm s}$, $\theta_{\rm s}$) and forward shock parameters ($\epsilon_{e}^{\rm fs}$, $\epsilon_{B}^{\rm fs}$, $f_e^{\rm fs}$). Our degrees of freedom for radio data fitting are consistent with previous works where $n_{\rm cnm},~k,~E_{\rm iso},~\theta_j,~s,~\Gamma,~\epsilon_e,~\epsilon_B$, and $\gamma_{e,\rm min}$ are typically needed to explain radio observations \citep[e.g.,][]{2023MNRAS.522.4028M}. The jet isotropic-equivalent energy $E_{\rm iso}$ encodes $\eta_{\rm s}$ and $\theta_{\rm s}$ which determine the energy deposited into the jet. Meanwhile, we {have used} $f_e$ to parameterize the electron minimum energy $\gamma_{e,\rm min}$. Therefore, we {have not introduced} extra degrees of freedom to fit the observations in one specific band.}

{{\bf Parameter degeneracy - }The fitting parameters summarized in Table \ref{tab:params} are obtained through theoretical estimations and manual adjustments. As discussed in the text, these parameters play a key role in determining the spectral/lightcurve shapes and flux levels. For instance, $\theta_{\rm f}$ influences both $L_{\rm f,\rm iso}$ and the jet break time through the beaming factor $\theta_{\rm f}^2/2$ and the jet break condition $\theta_{\rm f}\Gamma_{\rm f}=1$. The initial Lorentz factor $\Gamma_{\rm f0}$ determines the jet deceleration time and, consequently, the peak of the X-ray lightcurves. Additionally, $\alpha$ affects the slope of the X-ray lightcurve for $T_{\rm obs} \lesssim T_{\rm fb}$. Within the data fitting time window, $\Gamma_{\rm f}$ converges to the {behavior in the deceleration phase} and does not depend strongly on $\Gamma_{\rm f0}$. Therefore, we can roughly determine the values for $(\alpha,\eta_{\rm f},\Gamma_{\rm f0},\text{and}~\theta_{\rm f})$ from the X-ray lightcurve fitting. The remaining parameters ($s,\epsilon_e^{\rm rs},\epsilon_B^{\rm rs},~\text{and}~f_e^{\rm rs}$) primarily control the spectral shape. Specifically, $\gamma_{e,\rm m}^{\rm rs}$ is sensitive to $f_e^{\rm rs}$ rather than $\Gamma_{\rm f, rs-rel}$ and $s$, which, together with $\epsilon_B^{\rm rs}$, determine the synchrotron peak via $\nu_{\rm m} \propto (\gamma_{e,\rm m}^{\rm rs})^2 B_{\rm f,rs}$. Based on current observations, it is challenging to determine $f_e^{\rm rs}$ due to its degeneracy with $\epsilon_B^{\rm rs}$ and $\epsilon_e^{\rm rs}$, as studied in the context of GRBs \citep{2005ApJ...627..861E}. Nevertheless, a smaller $f_e^{\rm rs}<0.01$ is preferred to fit the X-ray spectra. Overall, the joint spectral and lightcurve fitting can reduce the degeneracy of the entire parameter space to some extent. We expect the degeneracy of slow jet parameters to be comparable with those of fast jet and previous work where analogous forward shock models are adopted \citep[e.g.,][]{2023MNRAS.522.4028M,2024ApJ...963...66Z} .}

\subsection{Multi-wavelength signatures}
{\bf X-ray variability} -- In addition to its rapid decay, another significant feature of the observed X-ray lightcurve is its rapid variability, characterized by a timescale of $\Delta T_{\rm var}\sim10^3$ s \citep{2024ApJ...965...39Y}. However, since we {have assumed} a homogeneous reverse shock downstream without considering the small scale plasma fluctuations, the intrinsic variability time scale of the reverse shock is much longer than $\Delta T_{\rm var}$
\begin{linenomath*}
\begin{equation}
    \Delta{T_{\rm var,rs}}=(1+z)\frac{R_{\rm f}}{\Gamma_{\rm f}^2\beta_{\rm f}c}\sim {T_{\rm obs}}\gg10^3~{\rm s}.
\end{equation}
\end{linenomath*}
This indicates that the fast variability cannot be attributed solely to the fast jet reverse shock region. In the reverse shock scenario, since the central engine is active, afterglow variabilities might arise when late-time outflow from the central engine collides with the preceding blast wave, resulting in a variability timescale down to the light crossing time of the central engine \citep{2005ApJ...631..429I}. Furthermore, the physical characteristics of the reverse shock region depend sensitively on $L_{\rm f,\rm iso}$, which can be modulated by accretion at the horizon of the SMBH \citep{2012Sci...337..949R}, leading to rapid variations on the timescale $\Delta T_{\rm var,eng}\sim (1+z)R_{\rm Sch}/c\sim10^3$ s, where $R_{\rm Sch}$ is the Schwarzschild radius. Overall, the continuously powered reverse shock scenario predicts both long-term (e.g., $\Delta T_{\rm var,rs}$) and rapid (e.g., $\Delta T_{\rm var,eng}$) X-ray variabilities, stemming from the reverse shock downstream and the central engine, respectively.

{\bf Late-time X-ray upper limits --} Recently, \cite{2024arXiv240410036E} reported the late-time X-ray upper limits in 0.3-10 keV band extending to $T_{\rm obs}\sim$400 d. These upper limits reveal a further steepened X-ray lightcurve, which was interpreted as the cessation of jet activity at $T_{\rm obs}\sim215$ d, when the accretion rate becomes sub-Eddington. In this paper, we {have demonstrated} that the steepening can also be alternatively explained by the jet break occurred at an earlier time, e.g., $T_{\rm br}\simeq72$ d. The late-time upper limits and the data points in the time interval $T_{\rm br}\lesssim T_{\rm obs}\lesssim 400$ d can be well described by the steepened lightcurve in the jet reverse shock scenario (see Fig. \ref{fig:lightcurve}).

{\bf Radio/millimeter emissions} -- We {have shown} that the radio spectra and lightcurves can be effectively described by the slow jet forward shock regions. However, the model-predicted early-time radio/millimeter energy fluxes fall below the measurements. This issue has also been noted in \cite{2024ApJ...963...66Z,2024arXiv240413326S}, where additional components or radiation zones are introduced to account for the early-time radio emissions. This suggests that the radio signals of TDEs may have a more complex origin.

{\bf $\mathbf{\gamma}$-ray detectability} -- Using the same parameters for the jet forward shock, we {have obtained} the energy flux $\nu F_\nu\sim10^{-14}-10^{-13}~\rm erg~s^{-1}~{cm^{-2}}$ in the \emph {Fermi} Large Area Telescope (LAT) $\gamma$-ray band (0.1$-$10 GeV). The non-detection of $\gamma$-ray sources by \emph{Fermi} within $1^\circ$ diameter of AT 2022cmc sets an upper limit on the energy flux $\nu F_{\nu}<8.8\times10^{-9}~\rm erg~s^{-1}~{cm^{-2}}$ \citep{2023NatAs...7...88P}, which is much higher than the $\gamma$-ray energy flux predicted by our wind-jet model. Since the forward shock lightcurve maintains a shallow decay before $T_{\rm br}$, we roughly estimate the 100-day detection horizon for AT 2022cmc-like TDEs for \emph{Fermi}-LAT in the optimistic case as $z\lesssim0.17$, This corresponds to the occurrence rate of approximately, $\lesssim0.02-0.1~\rm yr^{-1}$, using a rate density of $0.02^{+0.04}_{-0.01}~\rm Gpc^{-3}~yr^{-1}$ \citep{2022Natur.612..430A}. We {have inferred} that, it is difficult for \emph{Fermi} LAT to detect jetted TDE as luminous as AT 2022cmc within one decade, unless a much more efficient energy conversion in the forward shock region is attained.

{\bf Multimessenger implication} -- Since the identification of the first TDE with IceCube neutrino association, AT 2019dsg \citep{2021NatAs...5..510S}, more neutrino-coincident TDE candidates have been identified, such as AT 2019fdr \citep{2022PhRvL.128v1101R}, AT 2019aalc \citep{2021arXiv211109391V}, two dust-obscured ones \citep{2023ApJ...953L..12J}, and AT 2021lwx \citep{2024arXiv240109320Y}. Relativistic jets \citep{2016PhRvD..93h3005W,2017MNRAS.469.1354D,2017ApJ...838....3S,2017PhRvD..95l3001L}, sub-relativistic winds~\citep{2020ApJ...902..108M,2020PhRvD.102h3028L,2021NatAs...5..472W,2023ApJ...948...42W,2023ApJ...956...30Y}, and accretion flows~\citep{2020ApJ...902..108M,2021NatAs...5..436H} have been studied as the origin of the TDE neutrinos. For AT 2022cmc, we have tested the neutrino fluences using an efficient cosmic ray injection in the jet-wind model, where most of the jet power is converted to nonthermal protons. Our results demonstrate a very low neutrino fluence due to the relatively high redshift and less dense target photons, compared to AT 2019dsg, AT 2019fdr, and AT 2019aalc \citep[e.g.,][]{2020ApJ...902..108M,2023ApJ...948...42W,2023ApJ...956...30Y}. 

\section{Summary and conclusions}\label{sec:summary}

The distinct signatures of the radio/millimeter, optical/UV and X-ray signals of AT 2022cmc indicate a multi-component origin. In this work, we have presented a time-dependent structured jet model involving a fast relativistic jet (initial Lorentz factor of 30) and a slow relativistic wind (initial Lorentz factor of 4) to explain the multiwavelength spectral and temporal observations of AT 2022cmc. We {have modeled} the jet evolution within a generic external medium characterized by a power-law density profile, $n_{\rm ext}\propto R^{-k}$, extending to the ISM. Considering the active central engine powered by continuous accretion characterized by a power-law decaying accretion rate, we {have incorporated} persistent mass and power injections into the jets, which could significantly affect the dynamics and subsequently the multiwavelength lightcurves before the mass fallback time. For instance, continuous energy and mass injections extend the duration of the reverse shock emissions, enabling the explanation of late-time X-ray observations in the fast jet scenario.

We have self-consistently computed the synchrotron and inverse Compton emissions from the shock-accelerated energetic electrons in the forward and reverse shock regions of the fast and slow jets. We {have demonstrated} that the X-ray spectra and fast decaying lightcurves can be well described by the fast jet reverse shock region, whereas the slow jet forward shock could explain the radio/millimeter observations after $T_{\rm obs}\sim10$ d. Using the same forward shock parameters and reverse shocks, our calculation demonstrates that the fast jet forward shock and the slow jet reverse shock contributions can be subdominant. Specifically, we {have observed} that the jet forward shock would result in a more shallowly decaying lightcurve. Notably, the fast jet reverse shock X-ray lightcurve steepening due to the jet break at $T_{\rm br}\simeq72$ d aligns well with the late-time X-ray energy fluxes and upper limits extending to $T_{\rm obs}\sim400$ d. Using the same parameters for the jet forward shock region, we {have estimated} the $\gamma$-ray energy flux in the energy range 100 MeV to 10 GeV to be $\sim10^{-14}-10^{-13}~\rm erg~s^{-1}~cm^{-2}$ before $T_{\rm fb}$, which corresponds a detection rate of $\lesssim0.02-0.1$ per year for AT 2022cmc-like jetted TDEs and is consistent with the non-detection of jetted TDEs by \emph{Fermi} LAT.
The Klein-Nishina suppression to the inverse Compton emission together with the EBL absorption make it increasingly challenging to be detected in the very-high-energy TeV $\gamma$-ray ranges.

Our comprehensive modeling of the structured jet, involving the forward and reverse shock, related to TDEs provides a useful physical framework for interpreting the time-dependent multiwavelength observations of jetted TDEs detected in the future. Meanwhile, this work also provides a prototype to investigate the physical conditions of mass accretion, the ambient gas density profile, and the outflows, through spectral and lightcurve fitting.

\acknowledgements
We thank Mahmoud Al-Alawashra for the thorough review of the paper.
The work of K.M. is supported by the NSF Grant Nos.~AST-2108466, AST-2108467, and AST-2308021. B.T.Z. and K.M. are supported by KAKENHI Nos.~20H01901 and 20H05852.

\end{CJK*}
\appendix 


\section{Evolution of Relativistic Jets with Continuous Energy Injection: the full treatment with forward and reverse shocks}
\label{sec:app_jet}

\begin{figure*}
    \centering
    \includegraphics[width=0.7\textwidth]{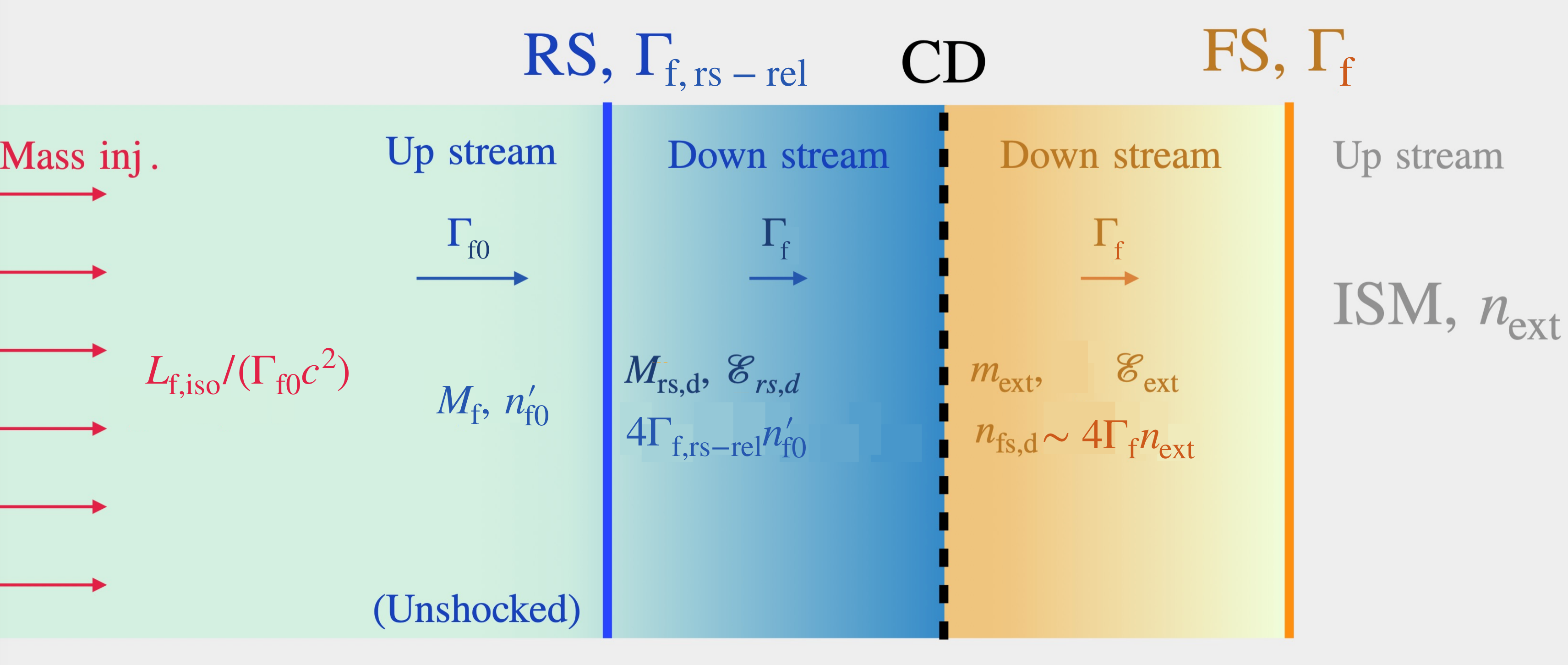}    \caption{Schematic figure of the fast jet with forward shock (FS), reverse shock (RS) and continuous energy injection. The contact discontinuity (CD) between the reverse shock and forward shock downstreams is also shown. The detailed description of the physical quantities can be found in the text.
}
    \label{fig:shock_schematic}
\end{figure*}

We analyze the evolution of a persistently powered fast jet within an external medium $n_{\rm ext}$, accounting for the effects of both forward and reverse shocks. Fig. \ref{fig:shock_schematic} schematically depicts the physical conditions of this jet, including the mass injections, contact discontinuity between the downstreams of reverse shock and forward shock, and the physical quantities (see the text below for the definitions) in different regions. Following the approach used in GRB afterglow modeling \citep{1999MNRAS.309..513H,2012ApJ...752L...8P,2013MNRAS.433.2107N,zhang2018physics,2023arXiv231113671Z}, we derive the total energy of the fast jet in the SMBH-rest frame,
\begin{equation}
    \mathcal E_{\rm f,\rm iso}=\Gamma_{\rm f0}M_{\rm f}c^2+\Gamma_{\rm f}M_{\rm rs,d}c^2+\Gamma_{\rm rs,eff}\mathcal E_{\rm rs,in}+\Gamma_{\rm f}m_{\rm ext}c^2+\Gamma_{\rm fs,eff}\mathcal E_{\rm ext,in},
\end{equation}
where $M_{\rm f}$, $M_{\rm rs,d}$ and $m_{\rm ext}$ are the masses of unshocked ejecta, reverse shock downstream, and the forward shock downstream originated from external medium. The downstream internal energies for forward shock and reverse shock can be written as
\begin{equation}
    \begin{split}
    \mathcal E_{\rm ext,in}&=(\Gamma_{\rm f}-1)m_{\rm ext}c^2,\\
    \mathcal E_{\rm rs,in}&=(\Gamma_{\rm f,rs-rel}-1)M_{\rm rs,d}c^2,
    \end{split}
\end{equation}
where $\Gamma_{\rm f, rs-rel}\approx(\Gamma_{\rm f}/\Gamma_{{\rm f}0}+\Gamma_{{\rm f}0}/\Gamma_{\rm f})/2$ is the relative Lorentz factor of the reverse shock and the adiabatic, and radiative energy losses are neglected \citep[see][for details]{zhang2018physics,2023arXiv231113671Z}. The effective Lorentz factor are
\begin{equation}
\begin{split}  
    \Gamma_{\rm fs,eff}&=(\hat\gamma_{\rm fs}\Gamma_{\rm f}^2-\hat\gamma_{\rm fs}+1)/\Gamma_{\rm f}\\
    \Gamma_{\rm rs,eff}&=(\hat\gamma_{\rm rs}\Gamma_{\rm f}^2-\hat\gamma_{\rm rs}+1)/\Gamma_{\rm {\rm f}},
    \end{split}
\end{equation}
where the adiabatic indices are defined as $\hat \gamma_{\rm fs}=(4\Gamma_{\rm f}+1)/(3\Gamma_{\rm f})$ and $\hat \gamma_{\rm rs}=(4\Gamma_{\rm f,rs-rel}+1)/(3\Gamma_{\rm f,rs-rel})$. 

Considering the energy injection from the SMBH and the swept-up external material, we have the derivative of $\mathcal E_{\rm \rm f,iso}$
\begin{equation}
    d\mathcal E_{\rm f,iso}=c^2dm_{\rm ext}+L_{\rm f,\rm iso}dt.
    \end{equation} 
We obtain the differential equations for $\Gamma_{\rm f}$ by combining the equations above,
\begin{equation}
    \frac{d\Gamma_{\rm f}}{dR_{\rm f}}=-\frac{\overbrace{(\Gamma_{\rm fs,eff}+1)(\Gamma_{\rm f}-1)c^2\frac{dm_{\rm ext}}{dR_{\rm f}}}^{\rm FS~term}+\overbrace{(\Gamma_{\rm f}-\Gamma_{{\rm f}0}-\Gamma_{\rm rs,eff}+\Gamma_{\rm rs,eff}\Gamma_{\rm f,rs-rel})c^2\frac{dM_{\rm rs,d}}{dR_{\rm f}}}^{\rm RS~term}}{(M_{\rm rs,d}+m_{\rm ext})c^2+\mathcal E_{\rm ext,in}\frac{d\Gamma_{\rm fs,eff}}{d\Gamma_{\rm f}}+\mathcal E_{\rm rs,in}\frac{d\Gamma_{\rm rs,eff}}{d\Gamma_{\rm f}}},
    \label{eq:jet_rs_fs}
\end{equation}
where the derivatives of $m_{\rm ext}$, $M_{\rm rs,d}$ and $M_{\rm f}$ can be explicitly written as
\begin{equation}
    \begin{split}
        \frac{dm_{\rm ext}}{dR_{\rm f}}&=4\pi R_{\rm f}^2n_{\rm ext}m_p,\\
        \frac{dM_{\rm rs,d}}{dR_{\rm f}}&=\dot M_{\rm rs,u}\left(\frac{dR_{\rm f}}{dt}\right)^{-1}\\&
        =\frac{3M_{\rm f}c}{R_{\rm f}}(\beta_{{\rm f}0}-\beta_{\rm f})\left(\frac{dR_{\rm f}}{dt}\right)^{-1},\\
    \frac{dM_{\rm f}}{dt}&=\frac{L_{{\rm f},\rm iso}}{\Gamma_{{\rm f}0}c^2}-\frac{3M_{\rm f}c}{R_{\rm f}}(\beta_{{\rm f}0}-\beta_{\rm f}).\\
      \end{split}
\end{equation}
In the above expressions, $\beta_{{\rm f}0}=\sqrt{1-\Gamma_{{\rm f}0}^{-2}}$ and $\beta_{{\rm f}}=\sqrt{1-\Gamma_{{\rm f}}^{-2}}$. Equation \ref{eq:jet_rs_fs} is consistent with Eq. A1 in \cite{2023arXiv231113671Z} and it reduces to the simple Eq. \ref{eq:jet_diff} if we neglect the influence of reverse shock by removing the differential terms of reverse shock and letting $M_{\rm rs,d}$ be the ejecta mass $M_{\rm ej}$. 

To get an intuition of the dominant factors, we observe that the reverse shock term is a high-order correction in Eq. \ref{eq:jet_rs_fs} since $(1-\Gamma_{\rm f}/\Gamma_{{\rm f}0})\rightarrow0$, $(\Gamma_{{\rm f,rs-rel}}-1)\rightarrow0$, and $(\beta_{{\rm f}0}-\beta_{\rm f})\rightarrow0$ when the jet is in the coasting regime. Later, in the deceleration stage we infer the ratio
\begin{equation}
    \zeta_{\rm rs/fs}=\frac{\rm RS~term}{\rm FS~term}< \frac{dM_{\rm rs,d}/{dR_{\rm f}}}{dm_{\rm ext}/{dR_{\rm f}}}\lesssim\frac{\bar n_{\rm rs}}{n_{\rm ext}}\frac{1}{2\Gamma_{\rm f}^2}\lesssim\frac{3\eta_{\rm f}\eta_{\rm acc}M_\star}{16\pi\theta_{\rm f}^2R_{\rm f}^3\Gamma_{{\rm f}0}\Gamma_{\rm f}^2m_pn_{\rm ext}},
    \label{eq:app:jet_diff}
\end{equation}
where $\bar n_{\rm rs}\sim M_{{\rm f}}/(4\pi m_pR_{\rm f}^3/3)$ is the average density of the reverse shock upstream, $M_{{\rm f}}\lesssim\int L_{{\rm f},\rm iso}dt/(\Gamma_{{\rm f}0}c^2)$, and the definition of $L_{\rm {\rm f},iso}$ are used. Plugging in the fiducial values, we find $\zeta_{\rm rs/fs}\lesssim10^{-2}$ for the fast and slow jet within the data-fitting time window, implying that the reverse shock (RS) term in the numerator is negligible. In the denominator, $M_{\rm rs,d}\approx M_{\rm ej}$ (ejecta mass in Eq. \ref{eq:jet_diff}) is satisfied as the reverse shock sufficiently crosses and decelerates the ejecta, thus reducing Eq. \ref{eq:jet_rs_fs} to Eq. \ref{eq:jet_diff}, and we conclude that the reverse shock does not significantly affect the jet evolution. Our calculation also suggests that the reverse shock term may play a role in jet evolution when the jet is exceptionally powerful (with higher $\eta_{\rm f}\eta_{\rm acc}M_\star$), extremely narrow (with smaller $\theta_{\rm f}$), and compact (with lower $\Gamma_{{\rm f}0}$ and $R_{\rm f}$).

\section{Analytical solutions}\label{sec:app_analytical}
Here, we take the fast jet as the example to derive the analytical Lorentz factor evolution, and the lightcurves of the forward shock and reverse shock scenarios.
\subsection{Jet Lorentz Factor}
We firstly derive the fast jet Lorentz factor as a function of $T_{\rm obs}$. Before the fallback time $T_{\rm fb}=(1+z)t_{\rm fb}$, we write down the isotropic equivalent energy of the jet
\begin{equation}
    \int L_{\rm f,\rm iso}dt \sim \frac{4\pi}{3}R_{\rm f}^2\Gamma_{\rm f}^3n_{\rm ext}m_pc^2.
\end{equation}
Noting $R_{\rm f}\sim \Gamma_{\rm f}^2 cT_{\rm obs}/(1+z)$, we obtain the time dependence of $\Gamma_{\rm f}$ in the deceleration regime 
\begin{equation}
    \Gamma_{\rm f}\propto T_{\rm obs}^{-(2+\alpha)/8},~{\rm for~}T_{\rm obs}<T_{\rm fb}.
\end{equation}
After $T_{\rm fb}$, $L_{{\rm f},\rm iso}$ decays faster than $t^{-1}$, which implies 
\begin{equation}
    \mathcal E_{{\rm f},\rm iso}\sim\frac{4\pi}{3}R_{\rm f}^3\Gamma_{\rm f}^2n_{\rm ext}m_pc^2 =\rm const,
\end{equation}
and 
\begin{equation}
\Gamma_{\rm f}\propto T_{\rm obs}^{-3/8},~{\rm for~}T_{\rm obs}>T_{\rm fb}. 
\end{equation}
The analytical solutions are consistent with Fig. \ref{fig:LorentzJ}. 

\subsection{Fast Jet Forward Shock}
Given the parameters in Table \ref{tab:params}, the injected electrons are in the fast cooling regime, i.e., the electron minimum Lorentz factor $\gamma_{e,m}^{\rm fs}$ is larger than the cooling Lorentz factor 
\begin{equation}
    \gamma_{e,c}^{\rm fs}=\frac{6\pi m_ec}{(1+Y)\sigma_TB_{\rm fs}^2t'_{\rm dyn}}
\end{equation}
where $Y\sim\mathcal O(\sqrt{\epsilon_e/\epsilon_B})$ is the Compton parameter \citep{2001ApJ...548..787S}, $\sigma_T$ is the Thomson cross section, and $t'_{\rm dyn}=R_{\rm f}/(\Gamma_{\rm f}c)$ is the comoving dynamic time. We estimate the peak synchrotron flux \citep[e.g.,][]{1999ApJ...523..177W}
\begin{equation}
    F_{\nu,\rm max}^{(\rm fs)}=\frac{(1+z)f_{\rm br}(0.6f_e^{\rm fs}n_{\rm ext}R_{\rm f}^3)\Gamma_{\rm f}e^3B_{\rm fs}}{\sqrt{3}m_ec^2d_L^2}\sim9.3\times10^{-27}{\rm erg~s^{-1}~cm^{-2}~Hz^{-1}}f_{\rm br}n_{\rm ext,0.3}^{3/2}\Gamma_{\rm f,0.8}^8(T_{\rm obs}/15~\rm d)^3.
\end{equation}
In the X-ray bands, e.g., $E_X=$10 keV$=h\nu_X$, we infer the energy flux
\begin{equation}
    \nu F_\nu^{\rm (fs)}=F_{\nu,\rm max}^{\rm (fs)}\nu_{c}^{1/2}\nu_{m}^{(s-1)/2}\nu_X^{(1-s/2)}\sim8.1\times10^{-14}{\rm erg~s^{-1}~cm^{-2}},
    \label{eq:app:fs_flux}
\end{equation}
where $\nu_c=3\Gamma_{\rm f}(\gamma_{e,c}^{\rm fs})^2eB/[4\pi (1+z)m_ec]$ is the cooling frequency.
The analytical value is consistent with the gray dashed curves in Fig. \ref{fig:specs}. Meanwhile, we derive the time dependence of the X-ray emissions produced by forward shock
\begin{equation}
    \nu F_\nu^{\rm (fs)}(E_X)\propto \begin{cases}
        f_{\rm br}T_{\rm obs}^{-\alpha+1-s/2},&T_{\rm obs}<T_{\rm br}\simeq T_{\rm fb}\\
        f_{\rm br}T_{\rm obs}^{-1}, & T_{\rm obs}>T_{\rm br}\simeq T_{\rm fb},
    \end{cases}
    \label{eq:app:FS_lightcurve}
\end{equation}
where $f_{\rm br}=1/[1+(\Gamma_{\rm f}\theta_{\rm f})^{-2}]$ is the jet break factor. In our calculation, we infer the jet break time, defined via $\Gamma_{\rm f}(t_{\rm br})=1/\theta_{\rm f}$, to be $t_{\rm br}\simeq2.8\times10^6~{\rm s}\simeq t_{\rm fb}$, which implies a steepened lightcurve for $T_{\rm obs}>T_{\rm br}\simeq T_{\rm fb}$,  $\nu F_\nu^{\rm (fs)}(E_X)\propto\Gamma_{\rm f}^2T_{\rm obs}^{-1}\propto T_{\rm obs}^{-7/4}$, noting that $\Gamma_{\rm f}\propto T_{\rm obs}^{-3/8}$.
Eq. \ref{eq:app:FS_lightcurve} indicates that the forward shock afterglow is disfavored to explain the fast decay of the X-ray lightcurve before and after the fallback time, since it predicts a subdominant slow decaying X-ray lightcurve.

\subsection{Fast Jet Reverse Shock}

In the deceleration regime, we have the relative Lorentz factor between the reverse shock up and downstreams $\Gamma_{\rm f, rs-rel}=(\Gamma_{\rm f0}/\Gamma_{\rm f}+\Gamma_{\rm f}/\Gamma_{\rm f0})/2$. The electrons accelerated in the reverse shock region are fast cooling using the parameters in Table \ref{tab:params}. Given the accelerated electron number in the reverse shock downstream, $N_{e,\rm rs}\sim f_e^{\rm rs}\Gamma_{\rm f,rs-rel}n_{\rm f0}'(4\pi R_{\rm f}^2)(R_{\rm f}/\Gamma_{\rm f})$, we write down the maximum synchrotron flux at $T_{\rm obs}=15$ d
\begin{equation}
    F_{\nu,\rm max}^{\rm rs}=\frac{0.6f_{\rm br}\sqrt{3}(1+z)N_{e,\rm rs}\Gamma_{\rm f}e^3B_{\rm f,rs}}{4\pi m_ec^2d_L^2}\simeq1.2\times10^{-26}{\rm erg~s^{-1}~cm^{-2}~Hz^{-1}}
\end{equation}
where fiducial values are used to calculate $L_{\rm f,\rm iso}$, $\Gamma_{\rm f,rs-rel}$ and $B_{\rm f,rs}$. From Fig. \ref{fig:specs}, we observe the relationship $\nu_X\gtrsim\nu_m>\nu_c$, where $\nu_m\propto \Gamma_{\rm f}(\gamma_{e,m}^{\rm rs})^2B_{\rm rs}$ is the characteristic frequency and $\nu_c$ is the cooling frequency defined similar to the forward shock case. In this case, we calculate the X-ray energy flux at $T_{\rm obs}=15$ d using the cooling frequency of the reverse shock regime,
\begin{equation}
        \nu F_\nu^{\rm (rs)}(E_X)=\nu_c^{1/2}\nu_m^{(s-1)/2}\nu_X^{(1-s/2)}F_{\nu,\rm max}^{\rm rs}\sim1.7\times10^{-12}{\rm erg~s^{-1}~cm^{-2}}.
\end{equation}
We also check the time evolution of X-ray lightcurves
\begin{equation}
    \nu F_\nu^{\rm (rs)}(E_X)\propto \nu_c^{1/2}\nu_m^{(s-1)/2}F_{\nu,\rm max}^{\rm rs}\propto\begin{cases}
        f_{\rm br}T_{\rm obs}^{-[5\alpha+\alpha(s-1)]/4}, &T_{\rm obs}<T_{\rm br}\simeq T_{\rm fb}\\
        f_{\rm br}T_{\rm obs}^{-(s+8)/6}, &T_{\rm obs}>T_{\rm br}\simeq T_{\rm fb}.\\
    \end{cases}
\end{equation}
Similar to the forward shock case, considering the jet break correction, we obtain a steepened lightcurve for $T_{\rm obs}>T_{\rm fb}\simeq T_{\rm br}$,  $\nu F_\nu^{\rm (rs)}(E_X)\propto\Gamma_{\rm f}^2T_{\rm obs}^{-(s+8)/6}\propto T_{\rm obs}^{-(2s+25)/12}\propto T_{\rm obs}^{-2.5}$ for $s=2.3$. Since we adopted the approximations, $\Gamma_{\rm f,rs-rel}\sim\Gamma_{\rm f, rs-rel}-1\sim \Gamma_{\rm f0}/(2\Gamma_{\rm f})$, to obtain Eq. \ref{eq:rs_lightcurves}, the numerical time decay slope could be steeper by a correction factor $\sim T_{\rm obs}^{-0.2}-T_{\rm obs}^{-0.4}$. In this case, the analytical lightcurve aligns well with numerical results depicted in Fig. \ref{fig:lightcurve} and is in agreement with late-time X-ray upper limits (the red trangles in Fig. \ref{fig:lightcurve}).
\bibliographystyle{aasjournal}
\bibliography{ref.bib}

\begin{thebibliography}{}
\expandafter\ifx\csname natexlab\endcsname\relax\def\natexlab#1{#1}\fi
\providecommand{\url}[1]{\href{#1}{#1}}
\providecommand{\dodoi}[1]{doi:~\href{http://doi.org/#1}{\nolinkurl{#1}}}
\providecommand{\doeprint}[1]{\href{http://ascl.net/#1}{\nolinkurl{http://ascl.net/#1}}}
\providecommand{\doarXiv}[1]{\href{https://arxiv.org/abs/#1}{\nolinkurl{https://arxiv.org/abs/#1}}}

\bibitem[{{Andreoni} {et~al.}(2022){Andreoni}, {Coughlin}, {Perley}, {Yao},
  {Lu}, {Cenko}, {Kumar}, {Anand}, {Ho}, {Kasliwal}, {de Ugarte Postigo},
  {Sagu{\'e}s-Carracedo}, {Schulze}, {Kann}, {Kulkarni}, {Sollerman}, {Tanvir},
  {Rest}, {Izzo}, {Somalwar}, {Kaplan}, {Ahumada}, {Anupama}, {Auchettl},
  {Barway}, {Bellm}, {Bhalerao}, {Bloom}, {Bremer}, {Bulla}, {Burns},
  {Campana}, {Chandra}, {Charalampopoulos}, {Cooke}, {D'Elia}, {Das}, {Dobie},
  {Ag{\"u}{\'\i} Fern{\'a}ndez}, {Freeburn}, {Fremling}, {Gezari}, {Goode},
  {Graham}, {Hammerstein}, {Karambelkar}, {Kilpatrick}, {Kool}, {Krips},
  {Laher}, {Leloudas}, {Levan}, {Lundquist}, {Mahabal}, {Medford}, {Miller},
  {M{\"o}ller}, {Mooley}, {Nayana}, {Nir}, {Pang}, {Paraskeva}, {Perley},
  {Petitpas}, {Pursiainen}, {Ravi}, {Ridden-Harper}, {Riddle}, {Rigault},
  {Rodriguez}, {Rusholme}, {Sharma}, {Smith}, {Stein}, {Th{\"o}ne},
  {Tohuvavohu}, {Valdes}, {van Roestel}, {Vergani}, {Wang}, \&
  {Zhang}}]{2022Natur.612..430A}
{Andreoni}, I., {Coughlin}, M.~W., {Perley}, D.~A., {et~al.} 2022, \nat, 612,
  430, \dodoi{10.1038/s41586-022-05465-8}

\bibitem[{{Barniol Duran} {et~al.}(2013){Barniol Duran}, {Nakar}, \&
  {Piran}}]{2013ApJ...772...78B}
{Barniol Duran}, R., {Nakar}, E., \& {Piran}, T. 2013, \apj, 772, 78,
  \dodoi{10.1088/0004-637X/772/1/78}

\bibitem[{{Berger} {et~al.}(2012){Berger}, {Zauderer}, {Pooley}, {Soderberg},
  {Sari}, {Brunthaler}, \& {Bietenholz}}]{2012ApJ...748...36B}
{Berger}, E., {Zauderer}, A., {Pooley}, G.~G., {et~al.} 2012, \apj, 748, 36,
  \dodoi{10.1088/0004-637X/748/1/36}

\bibitem[{{Blandford} \& {McKee}(1976)}]{1976PhFl...19.1130B}
{Blandford}, R.~D., \& {McKee}, C.~F. 1976, Physics of Fluids, 19, 1130,
  \dodoi{10.1063/1.861619}

\bibitem[{{Bloom} {et~al.}(2011){Bloom}, {Giannios}, {Metzger}, {Cenko},
  {Perley}, {Butler}, {Tanvir}, {Levan}, {O'Brien}, {Strubbe}, {De Colle},
  {Ramirez-Ruiz}, {Lee}, {Nayakshin}, {Quataert}, {King}, {Cucchiara},
  {Guillochon}, {Bower}, {Fruchter}, {Morgan}, \& {van der
  Horst}}]{2011Sci...333..203B}
{Bloom}, J.~S., {Giannios}, D., {Metzger}, B.~D., {et~al.} 2011, Science, 333,
  203, \dodoi{10.1126/science.1207150}

\bibitem[{{Brown} {et~al.}(2015){Brown}, {Levan}, {Stanway}, {Tanvir}, {Cenko},
  {Berger}, {Chornock}, \& {Cucchiaria}}]{2015MNRAS.452.4297B}
{Brown}, G.~C., {Levan}, A.~J., {Stanway}, E.~R., {et~al.} 2015, \mnras, 452,
  4297, \dodoi{10.1093/mnras/stv1520}

\bibitem[{{Burrows} {et~al.}(2011){Burrows}, {Kennea}, {Ghisellini}, {Mangano},
  {Zhang}, {Page}, {Eracleous}, {Romano}, {Sakamoto}, {Falcone}, {Osborne},
  {Campana}, {Beardmore}, {Breeveld}, {Chester}, {Corbet}, {Covino},
  {Cummings}, {D'Avanzo}, {D'Elia}, {Esposito}, {Evans}, {Fugazza}, {Gelbord},
  {Hiroi}, {Holland}, {Huang}, {Im}, {Israel}, {Jeon}, {Jeon}, {Jun}, {Kawai},
  {Kim}, {Krimm}, {Marshall}, {P. M{\'e}sz{\'a}ros}, {Negoro}, {Omodei},
  {Park}, {Perkins}, {Sugizaki}, {Sung}, {Tagliaferri}, {Troja}, {Ueda},
  {Urata}, {Usui}, {Antonelli}, {Barthelmy}, {Cusumano}, {Giommi}, {Melandri},
  {Perri}, {Racusin}, {Sbarufatti}, {Siegel}, \&
  {Gehrels}}]{2011Natur.476..421B}
{Burrows}, D.~N., {Kennea}, J.~A., {Ghisellini}, G., {et~al.} 2011, \nat, 476,
  421, \dodoi{10.1038/nature10374}

\bibitem[{{Cendes} {et~al.}(2021){Cendes}, {Eftekhari}, {Berger}, \&
  {Polisensky}}]{2021ApJ...908..125C}
{Cendes}, Y., {Eftekhari}, T., {Berger}, E., \& {Polisensky}, E. 2021, \apj,
  908, 125, \dodoi{10.3847/1538-4357/abd323}

\bibitem[{{Cenko} {et~al.}(2012){Cenko}, {Krimm}, {Horesh}, {Rau}, {Frail},
  {Kennea}, {Levan}, {Holland}, {Butler}, {Quimby}, {Bloom}, {Filippenko},
  {Gal-Yam}, {Greiner}, {Kulkarni}, {Ofek}, {Olivares E.}, {Schady},
  {Silverman}, {Tanvir}, \& {Xu}}]{2012ApJ...753...77C}
{Cenko}, S.~B., {Krimm}, H.~A., {Horesh}, A., {et~al.} 2012, \apj, 753, 77,
  \dodoi{10.1088/0004-637X/753/1/77}

\bibitem[{{Chevalier}(1998)}]{1998ApJ...499..810C}
{Chevalier}, R.~A. 1998, \apj, 499, 810, \dodoi{10.1086/305676}

\bibitem[{{Crumley} {et~al.}(2016){Crumley}, {Lu}, {Santana}, {Hern{\'a}ndez},
  {Kumar}, \& {Markoff}}]{2016MNRAS.460..396C}
{Crumley}, P., {Lu}, W., {Santana}, R., {et~al.} 2016, \mnras, 460, 396,
  \dodoi{10.1093/mnras/stw967}

\bibitem[{{Dai} \& {Fang}(2017)}]{2017MNRAS.469.1354D}
{Dai}, L., \& {Fang}, K. 2017, \mnras, 469, 1354, \dodoi{10.1093/mnras/stx863}

\bibitem[{{Eftekhari} {et~al.}(2018){Eftekhari}, {Berger}, {Zauderer},
  {Margutti}, \& {Alexander}}]{2018ApJ...854...86E}
{Eftekhari}, T., {Berger}, E., {Zauderer}, B.~A., {Margutti}, R., \&
  {Alexander}, K.~D. 2018, \apj, 854, 86, \dodoi{10.3847/1538-4357/aaa8e0}

\bibitem[{{Eftekhari} {et~al.}(2024){Eftekhari}, {Tchekhovskoy}, {Alexander},
  {Berger}, {Chornock}, {Laskar}, {Margutti}, {Yao}, {Cendes}, {Gomez},
  {Hajela}, \& {Pasham}}]{2024arXiv240410036E}
{Eftekhari}, T., {Tchekhovskoy}, A., {Alexander}, K.~D., {et~al.} 2024, arXiv
  e-prints, arXiv:2404.10036, \dodoi{10.48550/arXiv.2404.10036}

\bibitem[{{Eichler} \& {Waxman}(2005)}]{2005ApJ...627..861E}
{Eichler}, D., \& {Waxman}, E. 2005, \apj, 627, 861, \dodoi{10.1086/430596}

\bibitem[{{Evans} \& {Kochanek}(1989)}]{1989ApJ...346L..13E}
{Evans}, C.~R., \& {Kochanek}, C.~S. 1989, \apjl, 346, L13,
  \dodoi{10.1086/185567}

\bibitem[{Genet {et~al.}(2007)Genet, Daigne, \& Mochkovitch}]{Genet:2007nb}
Genet, F., Daigne, F., \& Mochkovitch, R. 2007, Mon. Not. Roy. Astron. Soc.,
  381, 732, \dodoi{10.1111/j.1365-2966.2007.12243.x}

\bibitem[{{Giannios} \& {Metzger}(2011)}]{2011MNRAS.416.2102G}
{Giannios}, D., \& {Metzger}, B.~D. 2011, \mnras, 416, 2102,
  \dodoi{10.1111/j.1365-2966.2011.19188.x}

\bibitem[{{Hayasaki}(2021)}]{2021NatAs...5..436H}
{Hayasaki}, K. 2021, Nature Astronomy, 5, 436,
  \dodoi{10.1038/s41550-021-01309-z}

\bibitem[{{Hills}(1975)}]{1975Natur.254..295H}
{Hills}, J.~G. 1975, \nat, 254, 295, \dodoi{10.1038/254295a0}

\bibitem[{{Huang} {et~al.}(1999){Huang}, {Dai}, \& {Lu}}]{1999MNRAS.309..513H}
{Huang}, Y.~F., {Dai}, Z.~G., \& {Lu}, T. 1999, \mnras, 309, 513,
  \dodoi{10.1046/j.1365-8711.1999.02887.x}

\bibitem[{{Ioka} {et~al.}(2005){Ioka}, {Kobayashi}, \&
  {Zhang}}]{2005ApJ...631..429I}
{Ioka}, K., {Kobayashi}, S., \& {Zhang}, B. 2005, \apj, 631, 429,
  \dodoi{10.1086/432567}

\bibitem[{{Jiang} {et~al.}(2023){Jiang}, {Zhou}, {Zhu}, {Wang}, \&
  {Wang}}]{2023ApJ...953L..12J}
{Jiang}, N., {Zhou}, Z., {Zhu}, J., {Wang}, Y., \& {Wang}, T. 2023, \apjl, 953,
  L12, \dodoi{10.3847/2041-8213/acebe3}

\bibitem[{{Kippenhahn} \& {Weigert}(1990)}]{1990sse..book.....K}
{Kippenhahn}, R., \& {Weigert}, A. 1990, {Stellar Structure and Evolution}

\bibitem[{{Klinger} {et~al.}(2023){Klinger}, {Rudolph}, {Rodrigues}, {Yuan},
  {Fichet de Clairfontaine}, {Fedynitch}, {Winter}, {Pohl}, \&
  {Gao}}]{2023arXiv231213371K}
{Klinger}, M., {Rudolph}, A., {Rodrigues}, X., {et~al.} 2023, arXiv e-prints,
  arXiv:2312.13371, \dodoi{10.48550/arXiv.2312.13371}

\bibitem[{{Liu} {et~al.}(2015){Liu}, {Pe'er}, \& {Loeb}}]{2015ApJ...798...13L}
{Liu}, D., {Pe'er}, A., \& {Loeb}, A. 2015, \apj, 798, 13,
  \dodoi{10.1088/0004-637X/798/1/13}

\bibitem[{{Liu} {et~al.}(2020){Liu}, {Xi}, \& {Wang}}]{2020PhRvD.102h3028L}
{Liu}, R.-Y., {Xi}, S.-Q., \& {Wang}, X.-Y. 2020, \prd, 102, 083028,
  \dodoi{10.1103/PhysRevD.102.083028}

\bibitem[{{Lunardini} \& {Winter}(2017)}]{2017PhRvD..95l3001L}
{Lunardini}, C., \& {Winter}, W. 2017, \prd, 95, 123001,
  \dodoi{10.1103/PhysRevD.95.123001}

\bibitem[{{Matsumoto} \& {Metzger}(2023)}]{2023MNRAS.522.4028M}
{Matsumoto}, T., \& {Metzger}, B.~D. 2023, \mnras, 522, 4028,
  \dodoi{10.1093/mnras/stad1182}

\bibitem[{{Metzger} {et~al.}(2012){Metzger}, {Giannios}, \&
  {Mimica}}]{2012MNRAS.420.3528M}
{Metzger}, B.~D., {Giannios}, D., \& {Mimica}, P. 2012, \mnras, 420, 3528,
  \dodoi{10.1111/j.1365-2966.2011.20273.x}

\bibitem[{{Mimica} {et~al.}(2015){Mimica}, {Giannios}, {Metzger}, \&
  {Aloy}}]{2015MNRAS.450.2824M}
{Mimica}, P., {Giannios}, D., {Metzger}, B.~D., \& {Aloy}, M.~A. 2015, \mnras,
  450, 2824, \dodoi{10.1093/mnras/stv825}

\bibitem[{{Murase} {et~al.}(2020){Murase}, {Kimura}, {Zhang}, {Oikonomou}, \&
  {Petropoulou}}]{2020ApJ...902..108M}
{Murase}, K., {Kimura}, S.~S., {Zhang}, B.~T., {Oikonomou}, F., \&
  {Petropoulou}, M. 2020, \apj, 902, 108, \dodoi{10.3847/1538-4357/abb3c0}

\bibitem[{{Murase} {et~al.}(2017){Murase}, {M{\'e}sz{\'a}ros}, \&
  {Fox}}]{2017ApJ...836L...6M}
{Murase}, K., {M{\'e}sz{\'a}ros}, P., \& {Fox}, D.~B. 2017, \apjl, 836, L6,
  \dodoi{10.3847/2041-8213/836/1/L6}

\bibitem[{{Nava} {et~al.}(2013){Nava}, {Sironi}, {Ghisellini}, {Celotti}, \&
  {Ghirlanda}}]{2013MNRAS.433.2107N}
{Nava}, L., {Sironi}, L., {Ghisellini}, G., {Celotti}, A., \& {Ghirlanda}, G.
  2013, \mnras, 433, 2107, \dodoi{10.1093/mnras/stt872}

\bibitem[{{Pasham} {et~al.}(2023){Pasham}, {Lucchini}, {Laskar}, {Gompertz},
  {Srivastav}, {Nicholl}, {Smartt}, {Miller-Jones}, {Alexander}, {Fender},
  {Smith}, {Fulton}, {Dewangan}, {Gendreau}, {Coughlin}, {Rhodes}, {Horesh},
  {van Velzen}, {Sfaradi}, {Guolo}, {Castro Segura}, {Aamer}, {Anderson},
  {Arcavi}, {Brennan}, {Chambers}, {Charalampopoulos}, {Chen}, {Clocchiatti},
  {de Boer}, {Dennefeld}, {Ferrara}, {Galbany}, {Gao}, {Gillanders}, {Goodwin},
  {Gromadzki}, {Huber}, {Jonker}, {Joshi}, {Kara}, {Killestein}, {Kosec},
  {Kocevski}, {Leloudas}, {Lin}, {Margutti}, {Mattila}, {Moore},
  {M{\"u}ller-Bravo}, {Ngeow}, {Oates}, {Onori}, {Pan}, {Perez-Torres}, {Rani},
  {Remillard}, {Ridley}, {Schulze}, {Sheng}, {Shingles}, {Smith}, {Steiner},
  {Wainscoat}, {Wevers}, \& {Yang}}]{2023NatAs...7...88P}
{Pasham}, D.~R., {Lucchini}, M., {Laskar}, T., {et~al.} 2023, Nature Astronomy,
  7, 88, \dodoi{10.1038/s41550-022-01820-x}

\bibitem[{{Pe'er}(2012)}]{2012ApJ...752L...8P}
{Pe'er}, A. 2012, \apjl, 752, L8, \dodoi{10.1088/2041-8205/752/1/L8}

\bibitem[{{Phinney}(1989)}]{1989IAUS..136..543P}
{Phinney}, E.~S. 1989, in The Center of the Galaxy, ed. M.~{Morris}, Vol. 136,
  543

\bibitem[{{Piran} {et~al.}(2015){Piran}, {Svirski}, {Krolik}, {Cheng}, \&
  {Shiokawa}}]{2015ApJ...806..164P}
{Piran}, T., {Svirski}, G., {Krolik}, J., {Cheng}, R.~M., \& {Shiokawa}, H.
  2015, \apj, 806, 164, \dodoi{10.1088/0004-637X/806/2/164}

\bibitem[{{Rees}(1988)}]{1988Natur.333..523R}
{Rees}, M.~J. 1988, \nat, 333, 523, \dodoi{10.1038/333523a0}

\bibitem[{{Reis} {et~al.}(2012){Reis}, {Miller}, {Reynolds}, {G{\"u}ltekin},
  {Maitra}, {King}, \& {Strohmayer}}]{2012Sci...337..949R}
{Reis}, R.~C., {Miller}, J.~M., {Reynolds}, M.~T., {et~al.} 2012, Science, 337,
  949, \dodoi{10.1126/science.1223940}

\bibitem[{{Reusch} {et~al.}(2022){Reusch}, {Stein}, {Kowalski}, {van Velzen},
  {Franckowiak}, {Lunardini}, {Murase}, {Winter}, {Miller-Jones}, {Kasliwal},
  {Gilfanov}, {Garrappa}, {Paliya}, {Ahumada}, {Anand}, {Barbarino}, {Bellm},
  {Brinnel}, {Buson}, {Cenko}, {Coughlin}, {De}, {Dekany}, {Frederick},
  {Gal-Yam}, {Gezari}, {Giroletti}, {Graham}, {Karambelkar}, {Kimura}, {Kong},
  {Kool}, {Laher}, {Medvedev}, {Necker}, {Nordin}, {Perley}, {Rigault},
  {Rusholme}, {Schulze}, {Schweyer}, {Singer}, {Sollerman}, {Strotjohann},
  {Sunyaev}, {van Santen}, {Walters}, {Zhang}, \&
  {Zimmerman}}]{2022PhRvL.128v1101R}
{Reusch}, S., {Stein}, R., {Kowalski}, M., {et~al.} 2022, \prl, 128, 221101,
  \dodoi{10.1103/PhysRevLett.128.221101}

\bibitem[{{Rhodes} {et~al.}(2023){Rhodes}, {Bright}, {Fender}, {Sfaradi},
  {Green}, {Horesh}, {Mooley}, {Pasham}, {Smartt}, {Titterington}, {van der
  Horst}, \& {Williams}}]{2023MNRAS.521..389R}
{Rhodes}, L., {Bright}, J.~S., {Fender}, R., {et~al.} 2023, \mnras, 521, 389,
  \dodoi{10.1093/mnras/stad344}

\bibitem[{{Rossi} {et~al.}(2002){Rossi}, {Lazzati}, \&
  {Rees}}]{2002MNRAS.332..945R}
{Rossi}, E., {Lazzati}, D., \& {Rees}, M.~J. 2002, \mnras, 332, 945,
  \dodoi{10.1046/j.1365-8711.2002.05363.x}

\bibitem[{{Sari} \& {Esin}(2001)}]{2001ApJ...548..787S}
{Sari}, R., \& {Esin}, A.~A. 2001, \apj, 548, 787, \dodoi{10.1086/319003}

\bibitem[{{Sato} {et~al.}(2024){Sato}, {Murase}, {Bhattacharya}, {Carpio},
  {Mukhopadhyay}, \& {Zhang}}]{2024arXiv240413326S}
{Sato}, Y., {Murase}, K., {Bhattacharya}, M., {et~al.} 2024, arXiv e-prints,
  arXiv:2404.13326, \dodoi{10.48550/arXiv.2404.13326}

\bibitem[{{Sato} {et~al.}(2021){Sato}, {Obayashi}, {Yamazaki}, {Murase}, \&
  {Ohira}}]{2021MNRAS.504.5647S}
{Sato}, Y., {Obayashi}, K., {Yamazaki}, R., {Murase}, K., \& {Ohira}, Y. 2021,
  \mnras, 504, 5647, \dodoi{10.1093/mnras/stab1273}

\bibitem[{{Senno} {et~al.}(2017){Senno}, {Murase}, \&
  {M{\'e}sz{\'a}ros}}]{2017ApJ...838....3S}
{Senno}, N., {Murase}, K., \& {M{\'e}sz{\'a}ros}, P. 2017, \apj, 838, 3,
  \dodoi{10.3847/1538-4357/aa6344}

\bibitem[{{Shen} \& {Matzner}(2014)}]{2014ApJ...784...87S}
{Shen}, R.-F., \& {Matzner}, C.~D. 2014, \apj, 784, 87,
  \dodoi{10.1088/0004-637X/784/2/87}

\bibitem[{{Stein} {et~al.}(2021){Stein}, {van Velzen}, {Kowalski},
  {Franckowiak}, {Gezari}, {Miller-Jones}, {Frederick}, {Sfaradi},
  {Bietenholz}, {Horesh}, {Fender}, {Garrappa}, {Ahumada}, {Andreoni},
  {Belicki}, {Bellm}, {B{\"o}ttcher}, {Brinnel}, {Burruss}, {Cenko},
  {Coughlin}, {Cunningham}, {Drake}, {Farrar}, {Feeney}, {Foley}, {Gal-Yam},
  {Golkhou}, {Goobar}, {Graham}, {Hammerstein}, {Helou}, {Hung}, {Kasliwal},
  {Kilpatrick}, {Kong}, {Kupfer}, {Laher}, {Mahabal}, {Masci}, {Necker},
  {Nordin}, {Perley}, {Rigault}, {Reusch}, {Rodriguez}, {Rojas-Bravo},
  {Rusholme}, {Shupe}, {Singer}, {Sollerman}, {Soumagnac}, {Stern}, {Taggart},
  {van Santen}, {Ward}, {Woudt}, \& {Yao}}]{2021NatAs...5..510S}
{Stein}, R., {van Velzen}, S., {Kowalski}, M., {et~al.} 2021, Nature Astronomy,
  5, 510, \dodoi{10.1038/s41550-020-01295-8}

\bibitem[{{Teboul} \& {Metzger}(2023)}]{2023ApJ...957L...9T}
{Teboul}, O., \& {Metzger}, B.~D. 2023, \apjl, 957, L9,
  \dodoi{10.3847/2041-8213/ad0037}

\bibitem[{{van Velzen} {et~al.}(2021){van Velzen}, {Stein}, {Gilfanov},
  {Kowalski}, {Hayasaki}, {Reusch}, {Yao}, {Garrappa}, {Franckowiak}, {Gezari},
  {Nordin}, {Fremling}, {Sharma}, {Yan}, {Kool}, {Sollerman}, {Medvedev},
  {Sunyaev}, {Bellm}, {Dekany}, {Duev}, {Graham}, {Kasliwal}, {Laher},
  {Riddle}, \& {Rusholme}}]{2021arXiv211109391V}
{van Velzen}, S., {Stein}, R., {Gilfanov}, M., {et~al.} 2021, arXiv e-prints,
  arXiv:2111.09391, \dodoi{10.48550/arXiv.2111.09391}

\bibitem[{{Wang} {et~al.}(2014){Wang}, {Lei}, {Wang}, {Zou}, {Zhang}, {Gao}, \&
  {Huang}}]{2014ApJ...788...32W}
{Wang}, J.-Z., {Lei}, W.-H., {Wang}, D.-X., {et~al.} 2014, \apj, 788, 32,
  \dodoi{10.1088/0004-637X/788/1/32}

\bibitem[{{Wang} \& {Liu}(2016)}]{2016PhRvD..93h3005W}
{Wang}, X.-Y., \& {Liu}, R.-Y. 2016, \prd, 93, 083005,
  \dodoi{10.1103/PhysRevD.93.083005}

\bibitem[{{Wijers} \& {Galama}(1999)}]{1999ApJ...523..177W}
{Wijers}, R.~A.~M.~J., \& {Galama}, T.~J. 1999, \apj, 523, 177,
  \dodoi{10.1086/307705}

\bibitem[{{Winter} \& {Lunardini}(2021)}]{2021NatAs...5..472W}
{Winter}, W., \& {Lunardini}, C. 2021, Nature Astronomy, 5, 472,
  \dodoi{10.1038/s41550-021-01305-3}

\bibitem[{{Winter} \& {Lunardini}(2023)}]{2023ApJ...948...42W}
---. 2023, \apj, 948, 42, \dodoi{10.3847/1538-4357/acbe9e}

\bibitem[{{Yao} {et~al.}(2024){Yao}, {Lu}, {Harrison}, {Kulkarni}, {Gezari},
  {Guolo}, {Cenko}, \& {Ho}}]{2024ApJ...965...39Y}
{Yao}, Y., {Lu}, W., {Harrison}, F., {et~al.} 2024, \apj, 965, 39,
  \dodoi{10.3847/1538-4357/ad2b6b}

\bibitem[{{Yuan} {et~al.}(2020){Yuan}, {Murase}, {Kimura}, \&
  {M{\'e}sz{\'a}ros}}]{2020PhRvD.102h3013Y}
{Yuan}, C., {Murase}, K., {Kimura}, S.~S., \& {M{\'e}sz{\'a}ros}, P. 2020,
  \prd, 102, 083013, \dodoi{10.1103/PhysRevD.102.083013}

\bibitem[{{Yuan} {et~al.}(2021){Yuan}, {Murase}, {Zhang}, {Kimura}, \&
  {M{\'e}sz{\'a}ros}}]{2021ApJ...911L..15Y}
{Yuan}, C., {Murase}, K., {Zhang}, B.~T., {Kimura}, S.~S., \&
  {M{\'e}sz{\'a}ros}, P. 2021, \apjl, 911, L15,
  \dodoi{10.3847/2041-8213/abee24}

\bibitem[{{Yuan} \& {Winter}(2023)}]{2023ApJ...956...30Y}
{Yuan}, C., \& {Winter}, W. 2023, \apj, 956, 30,
  \dodoi{10.3847/1538-4357/acf615}

\bibitem[{{Yuan} {et~al.}(2024){Yuan}, {Winter}, \&
  {Lunardini}}]{2024arXiv240109320Y}
{Yuan}, C., {Winter}, W., \& {Lunardini}, C. 2024, arXiv e-prints,
  arXiv:2401.09320, \dodoi{10.48550/arXiv.2401.09320}

\bibitem[{{Zauderer} {et~al.}(2013){Zauderer}, {Berger}, {Margutti}, {Pooley},
  {Sari}, {Soderberg}, {Brunthaler}, \& {Bietenholz}}]{2013ApJ...767..152Z}
{Zauderer}, B.~A., {Berger}, E., {Margutti}, R., {et~al.} 2013, \apj, 767, 152,
  \dodoi{10.1088/0004-637X/767/2/152}

\bibitem[{{Zauderer} {et~al.}(2011){Zauderer}, {Berger}, {Soderberg}, {Loeb},
  {Narayan}, {Frail}, {Petitpas}, {Brunthaler}, {Chornock}, {Carpenter},
  {Pooley}, {Mooley}, {Kulkarni}, {Margutti}, {Fox}, {Nakar}, {Patel},
  {Volgenau}, {Culverhouse}, {Bietenholz}, {Rupen}, {Max-Moerbeck}, {Readhead},
  {Richards}, {Shepherd}, {Storm}, \& {Hull}}]{2011Natur.476..425Z}
{Zauderer}, B.~A., {Berger}, E., {Soderberg}, A.~M., {et~al.} 2011, \nat, 476,
  425, \dodoi{10.1038/nature10366}

\bibitem[{Zhang(2018)}]{zhang2018physics}
Zhang, B. 2018, The physics of gamma-ray bursts (Cambridge University Press)

\bibitem[{{Zhang} \& {M{\'e}sz{\'a}ros}(2002)}]{2002ApJ...571..876Z}
{Zhang}, B., \& {M{\'e}sz{\'a}ros}, P. 2002, \apj, 571, 876,
  \dodoi{10.1086/339981}

\bibitem[{{Zhang} {et~al.}(2023){Zhang}, {Murase}, {Ioka}, \&
  {Zhang}}]{2023arXiv231113671Z}
{Zhang}, B.~T., {Murase}, K., {Ioka}, K., \& {Zhang}, B. 2023, arXiv e-prints,
  arXiv:2311.13671, \dodoi{10.48550/arXiv.2311.13671}

\bibitem[{{Zhang} {et~al.}(2021){Zhang}, {Murase}, {Veres}, \&
  {M{\'e}sz{\'a}ros}}]{2021ApJ...920...55Z}
{Zhang}, B.~T., {Murase}, K., {Veres}, P., \& {M{\'e}sz{\'a}ros}, P. 2021,
  \apj, 920, 55, \dodoi{10.3847/1538-4357/ac0cfc}

\bibitem[{{Zhou} {et~al.}(2024){Zhou}, {Zhu}, {Lei}, {Fu}, {Xie}, \&
  {Xu}}]{2024ApJ...963...66Z}
{Zhou}, C., {Zhu}, Z.-P., {Lei}, W.-H., {et~al.} 2024, \apj, 963, 66,
  \dodoi{10.3847/1538-4357/ad20f3}

\end{thebibliography}

\end{document}